\pdfoutput=1
\documentclass[11pt]{article}

\usepackage[letterpaper,margin=1in]{geometry}
\usepackage{natbib}
\renewcommand\cite{\citep}
\usepackage{caption}
\usepackage{times}
\usepackage{latexsym}
\usepackage{graphicx}
\usepackage{wrapfig}
\usepackage{booktabs}
\usepackage{array}
\usepackage{placeins}
\usepackage{multirow}
\usepackage{amsmath}
\usepackage{amssymb}
\usepackage{xcolor}

\usepackage{pifont}
\usepackage{tikz}
\usepackage{verbatim}
\usepackage{algorithm}
\usepackage[noend]{algpseudocode}
\usepackage[skins]{tcolorbox}
\usepackage{enumitem}
\definecolor{promptframe}{HTML}{1C5CAB}
\definecolor{promptback}{HTML}{F4F8FD}
\newtcolorbox{promptbox}[1]{enhanced, width=0.9\textwidth, center,
  colback=promptback, colframe=promptframe, coltitle=white,
  fonttitle=\sffamily\bfseries\normalsize, title={#1},
  arc=2mm, boxrule=0.6pt, left=3mm, right=3mm, top=1.5mm, bottom=1.5mm,
  fontupper=\sffamily\scriptsize}
\definecolor{dataframe}{HTML}{0E6E5C}
\definecolor{databack}{HTML}{F1F9F6}
\newtcolorbox{databox}[1]{enhanced, width=0.9\textwidth, center,
  colback=databack, colframe=dataframe, coltitle=white,
  segmentation style={draw=dataframe!45, line width=0.5pt},
  fonttitle=\sffamily\bfseries\normalsize, title={#1},
  arc=2mm, boxrule=0.6pt, left=3mm, right=3mm, top=1.5mm, bottom=1.5mm,
  fontupper=\sffamily\scriptsize}
\definecolor{caseframe}{HTML}{6E6587}
\definecolor{agentframe}{HTML}{5B8DB8}
\definecolor{agentback}{HTML}{EEF5FB}
\definecolor{envframe}{HTML}{6FA77B}
\definecolor{envback}{HTML}{F0F8F1}
\definecolor{casenote}{HTML}{8A6D3B}
\newtcolorbox{casebox}[1]{enhanced, colback=white, colframe=caseframe, coltitle=white,
  fonttitle=\sffamily\bfseries\normalsize, title={#1}, arc=2mm, boxrule=0.6pt,
  left=3mm, right=3mm, top=1.5mm, bottom=1.5mm, fontupper=\footnotesize}
\newtcolorbox{presbox}{enhanced, colback=caseframe!7, colframe=caseframe!25, boxrule=0.4pt,
  arc=1.5mm, borderline west={2pt}{0pt}{caseframe!80}, left=2.5mm, right=2.5mm,
  top=1mm, bottom=1mm, fontupper=\footnotesize}
\newtcolorbox{agentturn}{enhanced, width=0.84\linewidth, colback=agentback,
  colframe=agentframe!35, boxrule=0.4pt, arc=1.5mm, borderline west={2pt}{0pt}{agentframe},
  left=2.5mm, right=2.5mm, top=0.7mm, bottom=0.7mm, before skip=1pt, after skip=2pt, fontupper=\footnotesize}
\newtcolorbox{envturn}{enhanced, width=0.84\linewidth, flush right, colback=envback,
  colframe=envframe!35, boxrule=0.4pt, arc=1.5mm, borderline east={2pt}{0pt}{envframe},
  left=2.5mm, right=2.5mm, top=0.7mm, bottom=0.7mm, before skip=1pt, after skip=2pt, fontupper=\footnotesize}
\newcommand{\caseturn}[1]{\begin{center}\vspace{-2pt}\tcbox[on line, colback=caseframe,
  colframe=caseframe, coltext=white, arc=1mm, boxsep=0.4mm, left=2mm, right=2mm, top=0.3mm,
  bottom=0.3mm]{\sffamily\bfseries\scriptsize TURN #1}\vspace{-4pt}\end{center}}
\newcommand{\caserole}[2]{{\sffamily\bfseries\scriptsize\color{#1}#2}\par\smallskip}
\newcommand{\casenotes}[1]{\par\smallskip\hspace*{0.08\linewidth}\parbox{0.84\linewidth}{\sffamily
  \scriptsize{\color{casenote}\textbf{Note.}} #1}\par\medskip}
\newcommand{\pl}[1]{\textbf{#1}}
\newenvironment{pitems}{\begin{itemize}[leftmargin=1.3em, itemsep=0pt, parsep=0pt, topsep=1pt]}{\end{itemize}}

\usepackage{pifont}
\definecolor{markyes}{HTML}{3C8D40}
\definecolor{markno}{HTML}{E07A7A}
\definecolor{markprt}{HTML}{D4A72C}
\DeclareRobustCommand{\yes}{\textcolor{markyes}{\ding{51}}}
\DeclareRobustCommand{\no}{\textcolor{markno}{\ding{55}}}

\usepackage[T1]{fontenc}

\usepackage[utf8]{inputenc}

\usepackage{microtype}

\usepackage{inconsolata}

\usepackage{url}
\usepackage[breaklinks]{hyperref}
\definecolor{darkblue}{rgb}{0, 0, 0.5}
\hypersetup{colorlinks=true, citecolor=darkblue, linkcolor=darkblue, urlcolor=darkblue}

\title{Diagnosing Before the Answer: A Gated Environment for Evaluating LLM Agents on Real Emergency Chest-Pain Encounters}

\author{
  \textbf{Feng Wu\textsuperscript{1}},
  \textbf{Hongfei Yang\textsuperscript{1}},
  \textbf{Pengpeng Chen\textsuperscript{2}},
  \textbf{Kai He\textsuperscript{1}},
\\
  \textbf{Qika Lin\textsuperscript{1}},
  \textbf{Sicheng Hao\textsuperscript{3}},
  \textbf{Mengling Feng\textsuperscript{1}}
\\
  \small\textsuperscript{1}Saw Swee Hock School of Public Health, National University of Singapore
\\
  \small\textsuperscript{2}Sir Run Run Shaw Hospital, Zhejiang University School of Medicine
\\
  \small\textsuperscript{3}Alice Lee Centre for Nursing Studies, Yong Loo Lin School of Medicine, National University of Singapore
\\
  \small{\texttt{wufeng@u.nus.edu}, \texttt{hfyang@nus.edu.sg}, \texttt{3317054@zju.edu.cn}, \texttt{kai\_he@nus.edu.sg},}
\\
  \small{\texttt{linqika@nus.edu.sg}, \texttt{e1554787@u.nus.edu}, \texttt{mornin@nus.edu.sg}}
}

\date{}
\newlength{\introparindent}
\AtBeginDocument{\setlength{\introparindent}{\parindent}}

\begin{document}
\maketitle

\begin{abstract}
Large language models (LLMs) are increasingly evaluated on clinical diagnosis, but mostly on curated cases written after the answer was known, with closed label sets and accuracy as the only measure. Although this evaluation paradigm can measure LLM capabilities in a human-like way, it is far removed from clinical practice. To narrow this gap, we present EDBench-CP, a benchmark that turns 9{,}302 real emergency department chest-pain encounters from MIMIC-IV into an interactive environment, which contains beginning (history of present illness), process (examinations), and ending (diagnosis). In this environment, the agent needs to start from the triage record and arrival information, assess the patient's condition and order different tests, and based on the test results choose whether to continue testing or provide a diagnosis. To better reflect reality, diagnoses are given in free-text format, so we introduced LLM-as-a-judge, where the LLM scores the results according to a rubric calibrated by physicians. Across ten open-weight models, we find systematic gaps in choosing evidence, deciding when to stop, and recognising high-risk disease. Even the best agent is correct on only about half of the encounters, and every model misses 24\% to 45\% of time-critical diagnoses. The code and data will be released after review by PhysioNet.
\end{abstract}

\section{Introduction}
Large language models (LLMs) now pass medical licensing examinations~\citep{Nori2023,lievin2024can}, but an examination hands the model a complete vignette and a short list of answers. Diagnosis in the emergency department (ED) works differently. The physician starts from a chief complaint and a triage record, decides which examinations and tests to order and in what order, and decides when the evidence in hand is sufficient to finalise diagnoses. Measuring whether an LLM agent can do this requires an interactive evaluation environment rather than a question-answer set.

\begin{wrapfigure}{r}{0.45\textwidth}
  \vspace{-\baselineskip}
  \centering
  \includegraphics[width=\linewidth]{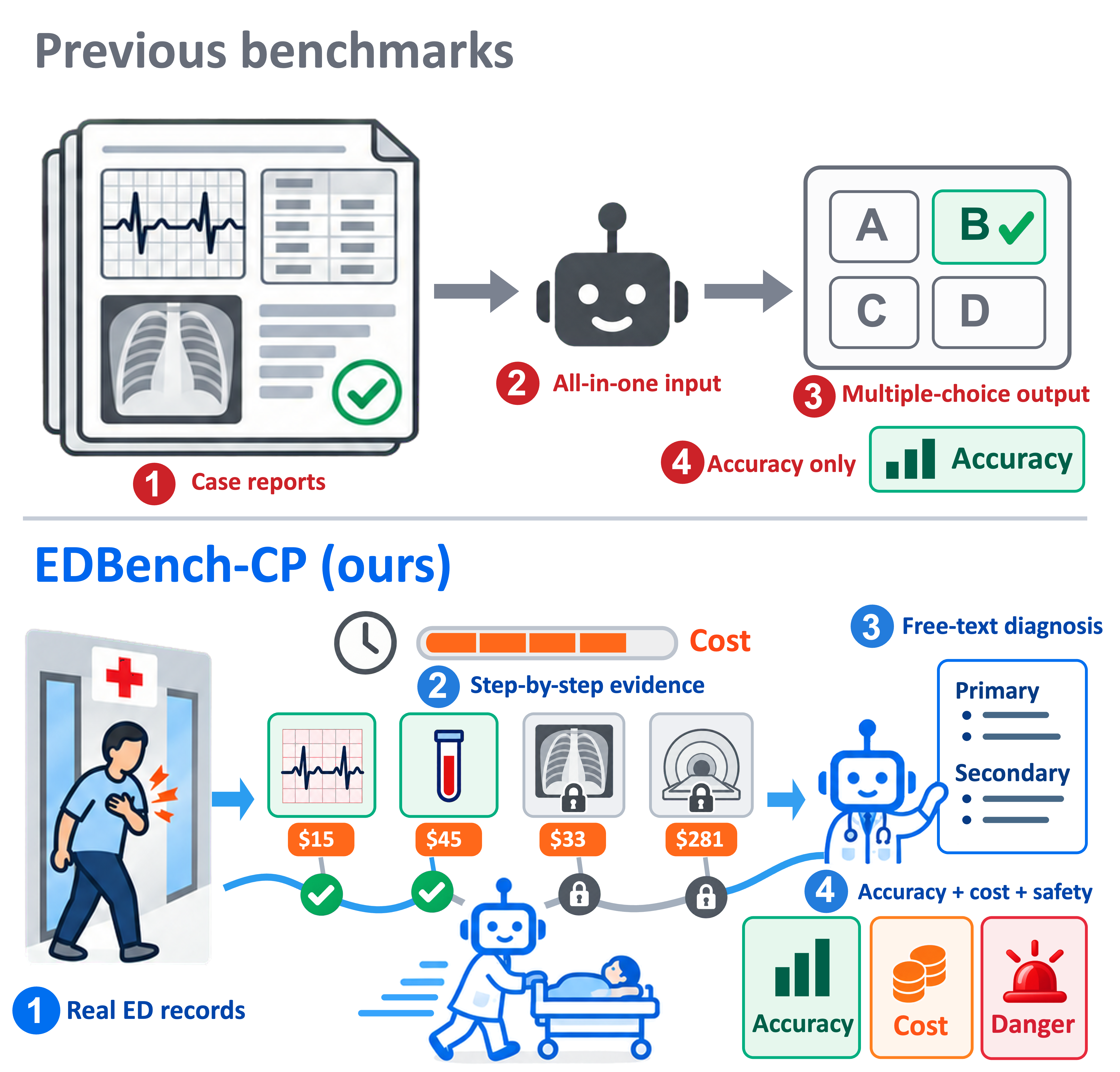}
  \caption{Previous benchmarks versus EDBench-CP on the four requirements of a diagnostic evaluation environment.}
  \label{fig:teaser}
  \vspace{-\baselineskip}
\end{wrapfigure}

We argue that such an environment should meet four requirements (Figure~\ref{fig:teaser}). \emph{Provenance}: cases should come from real encounters, not from case reports or teaching vignettes, as the former are usually more challenging due to missing information and conflicting evidence. \emph{Gated information}: evidence should be released only when the agent asks for it, so that the agent, not the benchmark, decides what to gather and when to stop. \emph{Open-ended target}: the agent should name the diagnosis in free text, because a single chief complaint has dozens of plausible diagnoses that no short candidate list can cover. \emph{Accuracy is not all}: the cost of the workup, and whether a time-critical diagnosis was missed, matter alongside accuracy. Existing benchmarks each relax at least one of these (Table~\ref{tab:relate work}). DiagnosisArena, VivaBench and SDBench build
{\parfillskip=0pt\par}
\noindent on case reports~\citep{zhu2026diagnosisarena,chiu2025vivabench,nori2025sequential}; DiReCT presents the whole record at once~\citep{wangdirect}; MIMIC-CDM gates real measurements but asks for one of four abdominal pathologies~\citep{hager2024evaluation}; and only SDBench prices the workup, while none reports missed time-critical diagnoses. To our knowledge, none satisfies all four.

We introduce EDBench-CP, an evaluation environment built from 9{,}302 real ED encounters in MIMIC-IV~\citep{PhysioNet-mimiciv-3.1} with chest pain symptom, one of the most common ED presentations, whose causes range from benign conditions such as musculoskeletal pain and gastro-oesophageal reflux to rapidly fatal ones such as myocardial infarction, pulmonary embolism and aortic dissection. Cases are not filtered by diagnosis, so the label space is long-tailed: 94 distinct primary diagnoses are needed to cover half of the encounters. The agent first sees only the triage record and the history available at arrival. All other evidence is gated behind tools for examination, ECG, laboratory tests and imaging; each call returns the patient's actual recorded result and is charged an estimated price, which the agent sees. The agent decides when it is ready and submits a free-text list of primary and secondary diagnoses, which a rubric-based LLM judge, calibrated against a clinician's ratings, grades against the discharge diagnosis. Our contributions are:
\begin{itemize}
\itemsep0em
    \item \textbf{An agent evaluation environment that meets all four requirements}.
    We are the first work to integrate these four requirements on a benchmark. This environment aligns with the real-world medical scenarios that doctors are experiencing rather than simple QA.
    Our dataset is also large enough to support training some models.
    \item \textbf{A clinician-calibrated LLM judge.} Calibrating the judge's rubric against a clinician's ratings reduces the judge's mean bias from $-0.32$ to $-0.03$ points, showing that an uncalibrated judge systematically under-scores agents.
    \item \textbf{A vanilla evaluation of ten open-weight LLMs and Physicians.} In our evaluation, we found that newer models perform better and can exceed physician accuracy while remaining cost-effective. However, the models still have problems such as (1) disobeying instructions, (2) repeatedly executing actions and (3) missing the critical clinical information.
\end{itemize}

\section{Related Works}

\subsection{Diagnostic Benchmarks for LLMs}
Diagnosis requires a physician to gather and integrate the chief complaint, history, examination findings and test results before committing to a decision~\citep{hager2024evaluation}. Table \ref{tab:relate work} summarizes previous benchmark work. These benchmarks still differ from emergency practice in several respects. First, most are built from curated material: DiagnosisArena, VivaBench and SDBench draw on published case reports~\citep{zhu2026diagnosisarena,chiu2025vivabench,nori2025sequential}, and MeDxAgent combines case reports with exam questions~\citep{sanghvi2026medxagent}. Such cases are written after the diagnosis is known and are selected for their teaching or publication value. Benchmarks built on clinical records avoid this, but DiReCT and MIMIC-CDM restrict the answer to a fixed set of diseases~\citep{wangdirect,hager2024evaluation}, and PhysicianBench contains only 100 cases~\citep{physicianbench2026}. Second, evaluation centers on accuracy. Only SDBench scores the cost of the tests a model orders and compares models with physicians, and none of these benchmarks reports missed time-critical diagnoses separately. Third, none of them provides non-text clinical data, and imaging is conveyed only as written descriptions. EDBench-CP addresses these gaps together, pairing 9{,}302 real ED encounters and their ECG images with open-ended grading, cost and safety evaluation, and a physician baseline.

\subsection{LLM Agents and Automated Evaluation in Clinical Diagnosis}
LLM agents for diagnosis increasingly divide the task among collaborating roles. MAI-DxO simulates a panel of physicians that maintains a differential diagnosis and selects cost-effective tests~\citep{nori2025sequential}, and MeDxAgent refines diagnostic hypotheses through a multi-agent consultation~\citep{sanghvi2026medxagent}. Because these agents produce open-ended diagnoses, their outputs are usually graded by another LLM: AI Hospital scores diagnostic reports with a GPT-4 chief physician~\citep{fan2025ai}, and MeDxAgent uses an LLM to decide whether a prediction and the reference name the same condition. EDBench-CP places no constraint on agent design, and its LLM judge compares each diagnosis list with the discharge diagnosis, with scores calibrated against a physician's ratings on our annotated subset.

{\setlength{\intextsep}{6pt}
\begin{table}[t]
\centering
\scriptsize
\begin{tabular}{llr ccccc}
\toprule
\textbf{Benchmark} & \textbf{Source} & \textbf{Cases} & \textbf{Open} & \textbf{Cost} & \textbf{Safety} & \textbf{Clin.} & \textbf{Multimodal}\\
\midrule
DiagnosisArena \cite{zhu2026diagnosisarena} & Case reports  & 1{,}113        & \yes & \no  & \no  & \no  & \no\\
VivaBench  \cite{chiu2025vivabench}    & Case reports     & 1{,}762        & \yes & \no  & \no  & \no  & \no\\
DiReCT \cite{wangdirect}        & MIMIC-IV & 511           & \no  & \no  & \no  & \no & \no \\
SDBench   \cite{nori2025sequential}     & Case reports      & 304            & \yes & \yes & \no  & \yes & \no\\
MIMIC-CDM  \cite{hager2024evaluation}    & MIMIC-IV      & 2{,}400        & \no  & \no  & \no  & \yes  & \no\\
MeDxAgent~\citep{sanghvi2026medxagent}       & Exams + case reports & 4{,}421 & \yes & \no  & \no  & \no  & \no\\
PhysicianBench \cite{physicianbench2026}& Clinical Records    & 100            & \yes & \no  & \no  & \no  & \no\\
\midrule
\textbf{EDBench-CP} & \textbf{MIMIC-IV} & \textbf{9{,}302} & \yes & \yes & \yes & \yes & \yes\\
\bottomrule
\end{tabular}
\caption{Comparison of existing diagnostic benchmarks. ``Open'', ``Cost'', ``Safety'', and ``Clin.'' stand for ``open-ended diagnosis'', ``test cost evaluation'', ``time-critical miss evaluation'', and ``clinician baseline'', respectively. ``Multimodal'' indicates whether the model receives non-text clinical data.}
\label{tab:relate work}
\end{table}
}

\section{EDBench-CP}

\begin{figure}[t]
    \centering
    \includegraphics[width=1\linewidth]{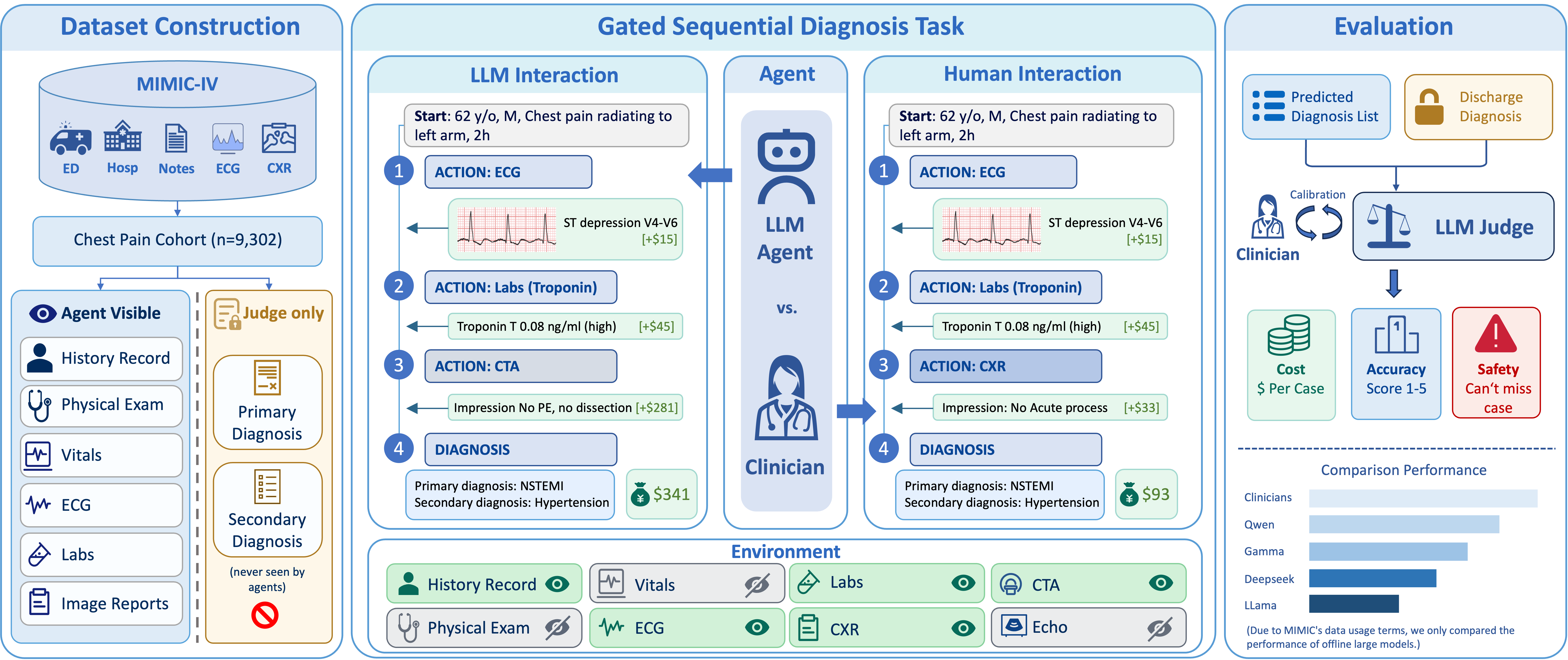}
    \caption{Overview of EDBench-CP.}
    \label{fig:overview}
\end{figure}

EDBench-CP converts real emergency department encounters into a sequential diagnosis task in which evidence is released only when requested (Figure~\ref{fig:overview}). Construction proceeds in three stages. We first select chest-pain encounters from MIMIC-IV and split each one into two parts: the evidence the treating team obtained during the workup, which the agent may request, and the discharge diagnosis, which only the judge sees. The encounter is then replayed as an interactive episode. The agent begins from the triage presentation, orders one test or examination at a time at a fixed price, and decides for itself when the evidence in hand justifies a diagnosis. Finally, the resulting diagnosis list is graded against the physician's free-text discharge diagnosis, and accuracy is reported together with workup cost and a dangerous-miss rate.

\subsection{Dataset Construction}

\paragraph{Cohort selection.}
We select ED encounters for chest pain from MIMIC-IV~\citep{PhysioNet-mimiciv-3.1}. An encounter is included when either the triage chief complaint or the chief complaint recorded in the discharge note indicates chest pain. The second criterion recovers 1{,}887 atypical presentations (20.3\% of the cohort), patients triaged for dyspnoea, palpitations or weakness whom a triage-only filter would miss. We exclude inter-facility transfers, encounters whose chief complaint is already a diagnosis (e.g.\ ``STEMI''), traumatic presentations, and encounters without an ECG during the ED stay. The final cohort contains 9{,}302 encounters. Appendix~\ref{app:cohort} gives the full criteria and the effect of each exclusion, and Appendix~\ref{app:evidence} describes how the evidence for each encounter is assembled.

\textbf{Why chest pain cohort}: We focus on chest pain for four reasons. First, its differential is unusually broad: it ranges from conditions that can be fatal within hours, such as myocardial infarction and aortic dissection, to benign ones such as musculoskeletal pain, which is exactly the setting the dangerous-miss rate is designed to measure. Second, its workup is defined by guidelines, with a clear escalation of tests as risk increases~\citep{gulati2021chestpain}, so it can test whether an agent orders investigations in proportion to the risk of the patient in front of it. Third, its diagnosis depends on change over time, such as a rising troponin or an evolving ECG, so the evidence has to be gathered over several steps rather than read off a single result. Finally, chest pain is one of the most common reasons for an emergency visit, which yields a large cohort: 9{,}302 encounters in MIMIC-IV.

\paragraph{Data collection.}
For each encounter we extract the triage record (demographics, chief complaint, acuity, pain score, arrival mode and vital signs), prior diagnoses, home medications and procedures, the admission physical examination, serial ED vital signs, every ECG with its machine-read report and waveform, laboratory results, and radiology reports, together with the discharge diagnosis that serves as the grading target. Evidence is taken from a window around the ED stay, below-detection laboratory values such as a troponin of ``$<$0.01'' are kept as negative results rather than dropped as missing, and echocardiography findings, which MIMIC-IV does not store in structured form, are parsed from the discharge note. The history of present illness in a discharge summary is written after the fact and often states the diagnosis, so we use an LLM to rewrite it as the history available at ED arrival and check the output for leaked results. Details are given in Appendix~\ref{app:evidence} and~\ref{app:hpi}.

\begin{figure}[t]
\begin{minipage}[t]{0.46\textwidth}
  \vspace{0pt}
  \centering
  \includegraphics[width=\linewidth]{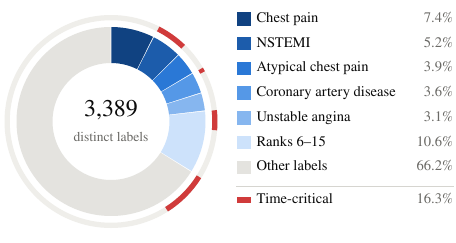}
  \caption{Distribution of primary discharge diagnoses.}
  \label{fig:labels}
\end{minipage}\hfill
\begin{minipage}[t]{0.5\textwidth}
  \vspace{0pt}
  \hrule height 0.8pt \kern 2pt
  \captionsetup{type=algorithm, skip=2pt, justification=raggedright, singlelinecheck=false}
  \caption{One episode of EDBench-CP.}
  \label{alg:episode}
  \kern 2pt \hrule \kern 3pt
\small
\begin{algorithmic}[1]
\Require encounter $(x_0, \mathcal{E}, y)$, agent $\pi$, step limit $T$
\State $h_1 \gets (x_0)$, \ $\mathcal{R}_0 \gets \emptyset$, \ $C \gets 0$
\For{$t = 1, \dots, T$}
  \If{$t = T$} require $\pi$ to diagnose \EndIf
  \State $(r_t, a_t) \gets \pi(h_t)$ \Comment{analysis, then action}
  \If{$a_t = \textsc{Diagnose}(\hat{y})$}
    \State \Return $\hat{y}$, $C$, $(\kappa_1, \dots, \kappa_{t-1})$
  \EndIf
  \State $o_t \gets g(a_t, \mathcal{R}_{t-1})$; \ $\mathcal{R}_t \gets \mathcal{R}_{t-1} \cup o_t$
  \State $C \gets C + c(a_t)$
  \State $\kappa_t \gets \mathbf{1}[\text{no item of } \mathcal{E} \text{ matches } a_t]$
  \State $h_{t+1} \gets (h_t, r_t, a_t, o_t, C)$
\EndFor
\end{algorithmic}
  \kern 2pt \hrule height 0.8pt
\end{minipage}
\end{figure}

\paragraph{Statistics.}
Figure~\ref{fig:labels} shows the distribution of the primary discharge diagnosis over the 9{,}073 encounters in which it could be parsed. The label space is long-tailed. Even after abbreviations and spelling variants are merged, the fifteen most frequent labels cover only 33.8\% of encounters and 94 labels are needed to cover half. The most frequent label, ``chest pain'', accounts for 7.4\%. A closed candidate list of any practical size would therefore exclude a large share of real answers.

\subsection{Gated Sequential Diagnosis Task} \label{sec:task}

\paragraph{Task formulation.}
Each encounter is a tuple $(x_0, \mathcal{E}, y)$. Here $x_0$ is the presentation at ED arrival (age, sex, chief complaint, triage acuity, pain score, mode of arrival, and the at-arrival history); $\mathcal{E}$ is the set of evidence items the treating team obtained, each with a type and a timestamp; and $y = (y^{\mathrm{p}}, Y^{\mathrm{s}})$ is the discharge diagnosis, consisting of a primary diagnosis and a set of secondary diagnoses, which the agent never observes. Starting from $x_0$, the agent must recover $y$ by requesting evidence from $\mathcal{E}$, and each request has a price.

\paragraph{Diagnostic process.}
At step $t$ the agent $\pi$ receives the interaction history $h_t = (x_0, r_1, a_1, o_1, \allowbreak \dots, \allowbreak r_{t-1}, a_{t-1}, o_{t-1})$ and produces $(r_t, a_t) = \pi(h_t)$ (Algorithm~\ref{alg:episode}). Each step combines three operations. \emph{Report analysis} ($r_t$) is a free-text interpretation of the evidence in hand, for example reading the ST segments of a returned ECG. It is written before the action, costs nothing and releases no evidence. \emph{Information gathering} requests one piece of evidence. The environment charges the price $c_t = c(a_t)$ and returns the report
\begin{equation}
o_t = g(a_t, \mathcal{R}_{t-1}), \qquad \mathcal{R}_t = \mathcal{R}_{t-1} \cup o_t,
\label{eq:step}
\end{equation}
where $\mathcal{R}_{t-1} \subseteq \mathcal{E}$ is the evidence released so far ($\mathcal{R}_0 = \emptyset$) and $g$ is the release function, which returns only unreleased items of $\mathcal{E}$ that match $a_t$. The report and the cost are appended to the history. \emph{Diagnosis} ($a_t = \textsc{Diagnose}$) submits $\hat{y} = (\hat{y}^{\mathrm{p}}, \hat{Y}^{\mathrm{s}})$ and ends the episode at $\tau = t$. Whether the evidence in hand justifies a diagnosis, or another test is worth its price, is left entirely to the agent.

Two safeguards complete the process. First, an agent can fall into a loop, for instance re-ordering a study whose results are exhausted, so the interaction is capped at $T = 20$ steps and the agent is required to diagnose at step $T$. Second, an agent may request a study that the treating team never performed, whose result is therefore unknown. Such a request is \emph{counterfactual}: the environment does not invent a result, returns a note that none is available, still charges $c_t$, and marks the step with $\kappa_t = 1$. The flags let us separate workups that stay within the recorded evidence from those that depart from it, and count how often an agent pays for tests that would not have been ordered in practice.

\subsection{Evaluation}
\paragraph{LLM-as-a-judge.} The same condition can be written in different words, such as ``NSTEMI'' and ``acute non-ST-elevation myocardial infarction'', while similar wording can hide a decisive difference like a type~A intramural haematoma calls for surgery, a descending aortic dissection usually for medical management. Lexical overlap therefore fails in both directions. Following \citet{nori2025sequential}, we instead grade with an LLM judge~\citep{zheng2023judging} that compares the submitted list with the discharge diagnosis on a five-point rubric written for this task. The benchmark fixes the rubric, not the judge model, so any sufficiently capable LLM can serve as the judge.

The rubric asks whether the submission names the cause of this chest-pain presentation. MIMIC-IV discharge summaries list every problem managed during the admission, so the judge first identifies the reference entry that explains the presentation and grades against it; unrelated in-hospital problems need not be listed. Scores of 5 and 4 name that cause as the primary diagnosis and differ only in the coverage of secondary diagnoses; 3 lists it only as secondary or confuses a sub-type; 2 gives a wrong but clinically adjacent primary diagnosis with similar ED management; and 1 gives one that changes management, or omits a time-critical diagnosis present in the reference. We count score $J \geq 4$ as correct, since management of the presenting complaint would remain largely unchanged above this threshold. The judge sees the reference diagnoses, with the ED diagnosis and the billing code as context, and the submitted list, which is always presented as an ED physician's. Grading is therefore blind to the transcript, to the cost, and to whether a model or a physician produced the diagnosis. During the evaluation process, the judge does not see the model's reasoning process or environment, so the score is based solely on the diagnosis. The complete scoring criteria and evaluation prompts are provided in Appendix~\ref{app:prompts}.

\paragraph{Metrics.}
We report four metrics in the evaluation. \emph{Diagnostic score} is the mean judge score $J$, together with accuracy, the share of encounters with $J \geq 4$. \emph{Cost} is the workup cost per encounter, $C = \sum_{t < \tau} c_t$ in US dollars, priced at the 2025/26 Medicare national average, e.g., flat \$45 per blood draw (the cost standard at Appendix~\ref{app:cost})~\citep{cms2026pfs,cms2026clfs}. \emph{Dangerous-miss rate} is the fraction with $J \leq 2$ among the encounters whose principal diagnosis is time-critical: MI, pulmonary embolism, aortic dissection, pneumothorax or cardiac tamponade. \emph{Interaction turns} is the mean episode length $\tau$, the number of steps up to and including the diagnosis, which shows how much evidence an agent gathers before committing.

\begin{table*}[!t]
\centering
\small
\begin{tabular*}{\textwidth}{@{\extracolsep{\fill}}lcccc@{}}
\toprule
Model & Score $\uparrow$ & Cost, median (IQR) $\downarrow$ & DMR $\downarrow$ & Turns, median (IQR) \\
\midrule
\multicolumn{5}{@{}l}{\textbf{Text-only input}} \\
\multicolumn{5}{@{}l}{\textit{General-purpose}} \\
\quad Qwen3.5-4B~\citep{qwen2026qwen35} & 2.82\,$\pm$\,1.51 & \$389 (314--650) & 36.0\% & 6 (5--8) \\
\quad Qwen3.5-9B~\citep{qwen2026qwen35} & 2.86\,$\pm$\,1.54 & \$419 (329--614) & 40.3\% & 6 (5--8) \\
\quad Qwen3.8-27B~\citep{qwen2026qwen38} & \textbf{3.33}\,$\pm$\,1.53 & \textbf{\$105} (60--308) & 30.2\% & 4 (4--5) \\
\quad Qwen3.6-35B-A3B~\citep{qwen2026qwen36} & 3.25\,$\pm$\,1.43 & \$374 (290--586) & \textbf{24.5\%} & 6 (5--8) \\
\quad Gemma-4-31B~\citep{gemma2026gemma4} & 2.98\,$\pm$\,1.53 & \$341 (138--740) & 35.3\% & 7 (5--9) \\
\quad Llama-4-Scout-17B-16E~\citep{meta2025llama4} & 2.97\,$\pm$\,1.45 & \$328 (290--538) & 39.6\% & 6 (5--6) \\
\quad gpt-oss-20b~\citep{openai2025gptoss} & 2.91\,$\pm$\,1.46 & \$425 (273--661) & 37.4\% & 7 (5--8) \\
\addlinespace[2pt]
\multicolumn{5}{@{}l}{\textit{Medical}} \\
\quad MedGemma-27B~\citep{sellergren2025medgemma} & 3.03\,$\pm$\,1.44 & \$332 (135--583) & 34.5\% & 7 (6--19) \\
\quad Hulu-Med-30A3~\citep{jiang2025hulumed} & 3.08\,$\pm$\,1.44 & \$420 (270--1063) & 25.9\% & 8 (6--20) \\
\quad Lingshu-32B~\citep{xu2025lingshu} & 2.73\,$\pm$\,1.43 & \$326 (118--604) & 45.3\% & 8 (7--9) \\
\midrule
\midrule
\multicolumn{5}{@{}l}{\textbf{Text + ECG image (multimodal models)}} \\
\multicolumn{5}{@{}l}{\textit{General-purpose}} \\
\quad Qwen3.5-4B~\citep{qwen2026qwen35} & 2.57\,$\pm$\,1.50 & \$464 (341--702) & 43.2\% & 6 (5--9) \\
\quad Qwen3.5-9B~\citep{qwen2026qwen35} & 2.70\,$\pm$\,1.48 & \$498 (341--831) & 41.7\% & 6 (5--10) \\
\quad Qwen3.8-27B~\citep{qwen2026qwen38} & \textbf{3.38}\,$\pm$\,1.47 & \textbf{\$105} (60--329) & 30.2\% & 4 (4--5) \\
\quad Qwen3.6-35B-A3B~\citep{qwen2026qwen36} & 3.19\,$\pm$\,1.49 & \$374 (257--586) & \textbf{24.5\%} & 7 (5--8) \\
\quad Gemma-4-31B~\citep{gemma2026gemma4} & 3.06\,$\pm$\,1.55 & \$364 (257--748) & 30.9\% & 7 (6--9) \\
\quad Llama-4-Scout-17B-16E~\citep{meta2025llama4} & 2.97\,$\pm$\,1.44 & \$300 (257--419) & 36.7\% & 5 (5--6) \\
\addlinespace[2pt]
\multicolumn{5}{@{}l}{\textit{Medical}} \\
\quad MedGemma-27B~\citep{sellergren2025medgemma} & 3.00\,$\pm$\,1.48 & \$374 (138--770) & 29.5\% & 9 (6--20) \\
\quad Hulu-Med-30A3~\citep{jiang2025hulumed} & 3.04\,$\pm$\,1.44 & \$372 (176--1036) & 27.3\% & 8 (6--20) \\
\quad Lingshu-32B~\citep{xu2025lingshu} & 2.68\,$\pm$\,1.65 & \$114 (93--341) & 48.9\% & 7 (6--8) \\
\bottomrule
\end{tabular*}
\caption{Main results in the gated setting on the 400-encounter subset. Score: judge score, mean\,$\pm$\,SD. Cost: workup cost in US dollars. DMR: dangerous-miss rate over the 139 time-critical encounters. Turns: episode length $\tau$. Best value per input setting in bold.}
\label{tab:main}
\end{table*}

\section{Experiments}
\subsection{Experimental Setup}
\paragraph{Backbone Models.}
We evaluate ten open-weight LLMs, all served locally with vLLM~\citep{kwon2023vllm}. Seven are general-purpose models: Qwen3.5-4B and Qwen3.5-9B~\citep{qwen2026qwen35}, Qwen3.6-35B-A3B~\citep{qwen2026qwen36}, Qwen3.8-27B~\citep{qwen2026qwen38}, Gemma-4-31B~\citep{gemma2026gemma4}, Llama-4-Scout-17B-16E (109B total, 17B active)~\citep{meta2025llama4} and gpt-oss-20b~\citep{openai2025gptoss}. Three are adapted to medicine: MedGemma-27B~\citep{sellergren2025medgemma}, Hulu-Med-30A3~\citep{jiang2025hulumed} and Lingshu-32B~\citep{xu2025lingshu}. The set spans 4B to 109B total parameters and includes both dense and mixture-of-experts architectures. All models except gpt-oss-20b accept images and are additionally evaluated with the rendered 12-lead ECG returned alongside its report. Every model acts in the identical environment with the same system prompt, action space, prices and step limit; serving and decoding settings are given in Appendix~\ref{app:setup}.

\textbf{Evaluation set and grading.}
We report results on a subset of 400 encounters, stratified by diagnostic category (19 categories; 139 encounters have a time-critical principal diagnosis), with physicians also completing the same task on the same subset. Diagnoses are graded by Claude Sonnet 5.5~\citep{anthropic2026sonnet55}.

\subsection{Results}

\paragraph{Main results.} Table~\ref{tab:main} reports the gated results, from which five findings emerge. \textit{No model solves the task}: even the strongest, Qwen3.8-27B, is correct on only about half of the encounters, most models cannot be statistically separated from their neighbours, and scale is not decisive, as a 27B dense model outperforms one with 109B total parameters. \textit{Accuracy does not require spending more}: with text-only input Qwen3.8-27B is also the cheapest agent, at under a third of the median cost of any other model, while the remaining models spend similar amounts for clearly different scores. \textit{Safety is not implied by accuracy}: every model misses at least a quarter of the time-critical encounters, and the dangerous-miss ranking departs from the score ranking, so a mean score alone would hide which models are safe. \textit{Medical adaptation does not help}: the medical models score no higher than the best general-purpose models of similar size and reach the step limit far more often, suggesting that domain training adds knowledge but not the judgement of when to stop gathering evidence. \textit{ECG images do not help}: adding the rendered 12-lead tracing leaves the score of nearly every multimodal model unchanged, including the medical models trained on medical images.

\begin{figure}[t]
  \centering
  \includegraphics[width=0.7\textwidth]{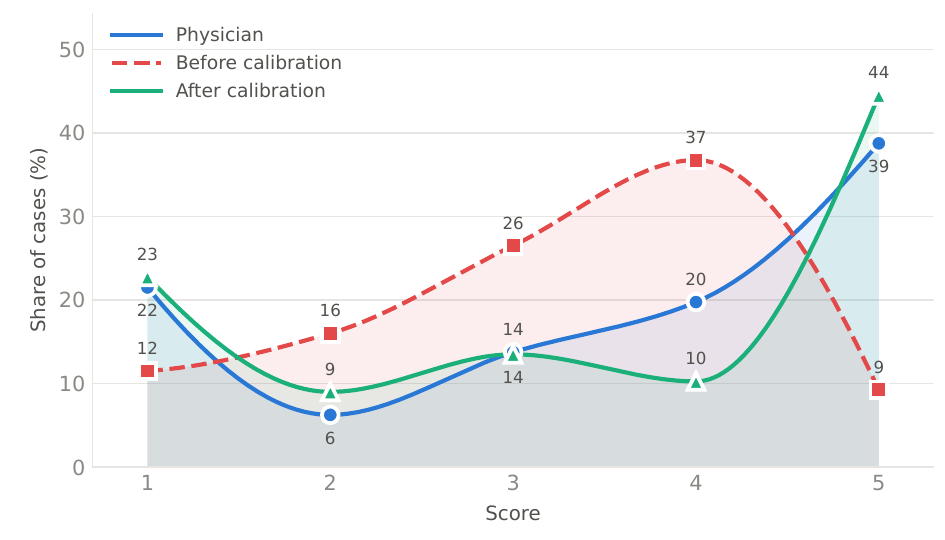}
  \caption{Score distribution graded by physician and by LLM judge before and after Calibration.}
  \label{fig:calibration}
\end{figure}

\paragraph{Judge calibration.}
Because grading relies on an LLM judge, its scores must agree with those of physicians. This calibration is performed by modifying the rubric. First, we had a physician rate the diagnoses generated by the model and the diagnoses in the ground truth. Based on the characteristics of the physician’s ratings, we adjusted our rubrics so that the LLM used for scoring would produce a similar distribution on the same diagnoses. The score distributions before and after calibration are shown in Figure~\ref{fig:calibration}. Before calibration, an earlier rubric and judge model were systematically stricter than the physician ($-0.32$ points on average) and rarely awarded full marks. With the final rubric, the judge matches the physician at both ends of the scale, the mean bias falls to $-0.03$. After obtaining a rubric aligned with physicians, we re-scored the results of all models to obtain more accurate accuracy scores. The calibrated rubric is shown in the Appendix~\ref{app:judge}.

\begin{figure}[t]
  \centering
  \includegraphics[width=0.7\textwidth]{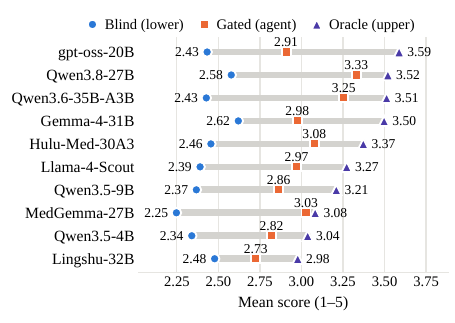}
  \caption{Gated score of each model between its blind lower bound (presentation only) and oracle upper bound (all evidence at once).}
  \label{fig:anchors}
\end{figure}

\paragraph{Lower and upper bounds.}
A gated score conflates deciding which evidence to acquire with interpreting it once it arrives, so a weak agent may be one that ordered the wrong tests or one that ordered the right tests and failed to read them. To separate the two we bracket each gated run with single-turn controls on the same encounters, graded by the same rubric and judge: in the \emph{blind} setting the model sees only the presentation $x_0$ and answers immediately; in the \emph{oracle} setting it receives the entire recorded workup $\mathcal{E}$ at once. Figure~\ref{fig:anchors} places each gated score in the resulting interval, whose width $\text{oracle} - \text{blind}$ is the value of the evidence and within which $(\text{gated} - \text{blind}) / (\text{oracle} - \text{blind})$ is the fraction the agent realises. Given only the HPI, all ten models score within a narrow band, so prior knowledge barely separates them. Given all the evidence at once the spread roughly doubles, so it is the integration of evidence rather than prior knowledge that distinguishes models, and every model gains substantially from the full record, confirming that the recorded workup carries diagnostic information. Gated LLMs, however, recover only part of this gain, most realising between two fifths and four fifths of it, so deciding what to order costs them a large share of what the evidence could provide. Qwen3.8-27B leads the gated setting because it combines a high upper bound with one of the highest realisations, whereas Gemma-4-31B and gpt-oss-20b reach upper bounds comparable to the best models but lose most of that advantage when they must choose their own tests.
Further analyses, including cost--accuracy and budget curves and per-model score distributions, are given in Appendix~\ref{app:results}.

\section{Discussion}
\label{sec:discussion}

\begin{table}[t]
\centering
\small
\setlength{\tabcolsep}{2.5pt}
\begin{tabular}{@{}lcccc@{}}
\toprule
 & Score $\uparrow$ & Cost $\downarrow$ & DMR $\downarrow$ & Turns \\
\midrule
\multicolumn{5}{@{}l}{\textit{General-purpose}} \\
\quad Qwen3.5-4B & 2.51\,$\pm$\,1.49 & \$461 & 46.4\% & 6 (5--9) \\
\quad Qwen3.5-9B & 2.69\,$\pm$\,1.44 & \$479 & 39.2\% & 7 (5--9) \\
\quad Qwen3.8-27B & \textbf{3.41}\,$\pm$\,1.43 & \textbf{\$105} & 24.8\% & 4 (4--5) \\
\quad Qwen3.6-35B-A3B & 3.21\,$\pm$\,1.50 & \$379 & \textbf{22.4\%} & 7 (5--8) \\
\quad Gemma-4-31B & 3.05\,$\pm$\,1.53 & \$343 & 27.2\% & 7 (6--9) \\
\quad Llama-4-Scout & 2.96\,$\pm$\,1.45 & \$310 & 34.4\% & 5 (5--6) \\
\midrule
\multicolumn{5}{@{}l}{\textit{Medical}} \\
\quad MedGemma-27B & 2.98\,$\pm$\,1.50 & \$355 & 31.2\% & 8 (6--20) \\
\quad Hulu-Med-30A3 & 3.04\,$\pm$\,1.44 & \$374 & 23.2\% & 8 (6--20) \\
\quad Lingshu-32B & 2.68\,$\pm$\,1.63 & \$123 & 49.6\% & 7 (7--8) \\
\midrule
\textbf{Physicians} & 2.70\,$\pm$\,1.54 & \$613 & 33.6\% & 8 (7--10) \\
\bottomrule
\end{tabular}
\caption{The performance of doctors and VLM models on the cleaned validation set (307 cases).}
\label{tab:physician}
\end{table}

\begin{figure*}[t]
  \centering
  \includegraphics[width=\textwidth]{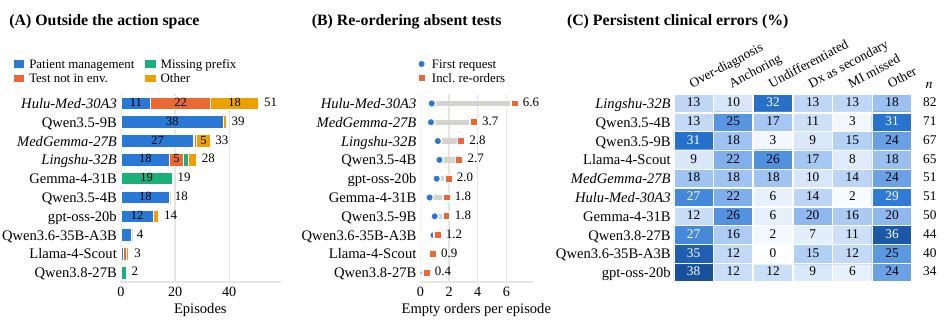}
  \caption{(A)~Requests outside the action space, counted once per episode. (B)~Orders returning no result, per episode: first request for an absent test, and the total including repeats. (C)~Clinical error modes of failures that persist with all evidence (gated and oracle score $\leq 2$), as a share of each model's persistent failures.}
  \label{fig:failures}
\end{figure*}

\paragraph{Can LLMs Diagnose Like Doctors?}

We provided the same validation set to human physicians for evaluation. We recruited four emergency department physicians in China to make diagnoses using an annotation tool (see Appendix~\ref{app:annotation}). To adapt to China’s medical environment, we translated the cases into Chinese and made some localization adjustments. After receiving the doctors’ results, we removed some low-quality annotations, including submissions with 0 examination and submissions completed in less than 30 seconds. We obtained 307 annotated cases, and the results are shown in Table~\ref{tab:physician}. Physicians achieved a score of 2.7, but their DMR of 33.6\% was lower than that of four models (two of which were medical models), while physicians had the highest costs. This stems from characteristics of the Chinese medical environment: physicians tend to obtain more examination results to make decisions (defensive medicine). In addition, these numbers describe the performance of four physicians in an unfamiliar environment; the range of dangerous missed diagnoses among the four annotators (25.6\% to 50.0\%) was even larger than that among models (see Appendix). Therefore, the physicians’ ratings were constrained by the physicians’ level of expertise.

\paragraph{Failure Mode Analysis.} Figure~\ref{fig:failures} separates three kinds of failure: two concern how agents use the environment, and one concerns how they reason about the evidence.
(1). \textbf{Acting outside the action space} (Figure~\ref{fig:failures}A). In 211 of the 4{,}000 text-only episodes the agent issues a request the environment cannot execute. Most are management rather than investigation, such as waiting for a repeat measurement, calling a consultant or giving treatment. Others name tests the environment does not offer, chiefly coronary angiography and stress testing, and the rest are malformed orders. The LLMs behave as if they were managing the patient, not as if they lacked medical knowledge.
(2). \textbf{Re-ordering a test the encounter does not contain} (Figure~\ref{fig:failures}B). Although the environment states that no result is available, several models order the same test again until the step limit ends the episode. The medical models do this most, like Hulu-Med-30A3, when it reaches the step limit, two thirds of its steps re-order an absent test, most often CT angiography. This indicates that the model cannot understand the current patient status and cannot provide a good diagnosis.
(3). \textbf{Clinical reasoning issues} (Figure~\ref{fig:failures}C), and the failures share one habit: a single test value decides the diagnosis. For example, in failed infarction cases, agents see about one troponin when three are available. Conversely, a negative troponin is taken to exclude a cardiac cause, so unstable angina in patients with known coronary disease is called non-cardiac chest pain, and pulmonary embolism and aortic dissection are missed because CTA is never ordered once ACS seems excluded. This is not only a failure to gather evidence: in 45\% of failed gated episodes, the same model also fails when given the whole record at once. These persistent errors are led by over-diagnosis and by anchoring on heart failure, a laboratory abnormality or another organ system, which together account for about 40\%, followed by a clear cause called undifferentiated chest pain. In each, one salient finding outweighs the history and the trend, whereas emergency physicians set the prior from the history and update on trends.

We also investigate  \textbf{unavailable-order rate}, \textbf{evidence utilization} and \textbf{cost sensitivity} in Appendix ~\ref{app:results}. And the success and failed case study could be found in Appendix ~\ref{app:case}.

\section{Conclusion}
We proposed EDBench-CP, a benchmark for evaluating LLM agents on sequential diagnosis of real emergency chest-pain encounters, in which evidence must be requested at a cost and diagnoses are graded against free-text discharge diagnoses. Our evaluations show that even the strongest of ten open-weight models is correct on only about half of the encounters and that every model misses at least a quarter of time-critical diagnoses, underscoring the need for agents that know what to order and when the evidence suffices. Because the environment depends only on the case schema, EDBench-CP extends to other chief complaints and data sources with a new extractor, facilitating the evaluation of future clinical agents.

\section{Limitations}
EDBench-CP trades the control of curated cases for the realism of recorded encounters, and most of its limitations follow from that choice. The environment can only replay the workup the treating team actually ordered: an agent that takes a different but reasonable path, such as a repeat troponin the team never requested, is told that no result exists and is still charged, so alternative routes cannot be credited, and the reply itself reveals what the team did not order. Some encounters cannot be solved from their record at all, because decisive evidence lies outside it (outside-hospital care, images seen only by the radiologist, unrecorded bedside findings), the admission examination may postdate the ED stay, and the discharge diagnosis used as the reference can rest on later information or contain errors.

The benchmark is also narrow in scope. All encounters come from one academic medical centre in Boston and present with chest pain, and costs use US Medicare prices with fixed turnaround times, so absolute costs do not transfer to other health systems. The agent receives the history as one block instead of eliciting it, and it neither treats the patient nor decides disposition, although real ED diagnosis is interleaved with management.

Comparing LLMs with physicians is a direction we plan to pursue in depth. We will run a multi-centre retrospective study in which physicians from several hospitals work up the same encounters in the same environment, each in their own language and documentation conventions, so that physicians and LLMs can be compared case by case. The physician results in this paper are a first step towards that study and serve as a reference point: the physicians read machine-translated cases and wrote their diagnoses in Chinese, and each worked up a different set of encounters, so their scores carry a shift from translation and case mix.

In our future work, we will introduce external validation datasets to assess LLMs’ understanding of healthcare systems and doctor-patient situations across multiple centers and countries. We will also enhance the coverage of examinations and the corresponding management of certain emergency conditions, so as to better simulate real-world scenarios.

\bibliography{custom}

\clearpage
\appendix

\renewcommand{\topfraction}{0.9}
\renewcommand{\dbltopfraction}{0.9}
\renewcommand{\textfraction}{0.07}
\renewcommand{\floatpagefraction}{0.7}
\renewcommand{\dblfloatpagefraction}{0.7}
\setcounter{topnumber}{3}
\setcounter{dbltopnumber}{3}
\setcounter{totalnumber}{5}

\section{Ethics Statement} \label{app:ethics}

\paragraph{Data access.} EDBench-CP is derived from MIMIC-IV~\citep{PhysioNet-mimiciv-3.1}, which is de-identified under the HIPAA Safe Harbor provision before release and distributed by PhysioNet under a credentialed data use agreement. Everyone who handled the records, including the physicians who annotated, completed the required training and signed the agreement, and we made no attempt to re-identify any patient. The collection of MIMIC-IV was approved by the institutional review boards of Beth Israel Deaconess Medical Center and the Massachusetts Institute of Technology, which waived the requirement for informed consent.

\paragraph{Data release.} We will release EDBench-CP on PhysioNet as a derived dataset under the same credentialed access and data use agreement as MIMIC-IV, together with the code for the environment and the grading, so that access to the patient records remains governed by PhysioNet.

\paragraph{Human participants.} Four attending emergency physicians took part in the physician evaluation (Appendix~\ref{app:physicians}) and one in the calibration of the judge (Appendix~\ref{app:judge}). They were recruited from Sir Run Run Shaw Hospital, Hangzhou, China, were informed of the purpose of the study and of how their diagnoses would be used, and agreed to take part; their results are reported anonymously.

\paragraph{Local processing.} All evaluated agents and the rewriting of the history of present illness ran on locally served models.

\paragraph{Intended use.} EDBench-CP is a research benchmark, not a clinical tool. The models we evaluate miss at least a quarter of time-critical diagnoses and should not inform the care of real patients. The cohort comes from a single academic centre and contains only admitted patients, so results may not transfer to other populations or settings; we include atypical presentations, which are more common in women, diabetic and elderly patients, to reduce one known source of bias.

\FloatBarrier
\section{Dataset Construction Details} \label{app:data}

\begin{figure*}[t]
  \centering
  \includegraphics[width=\textwidth]{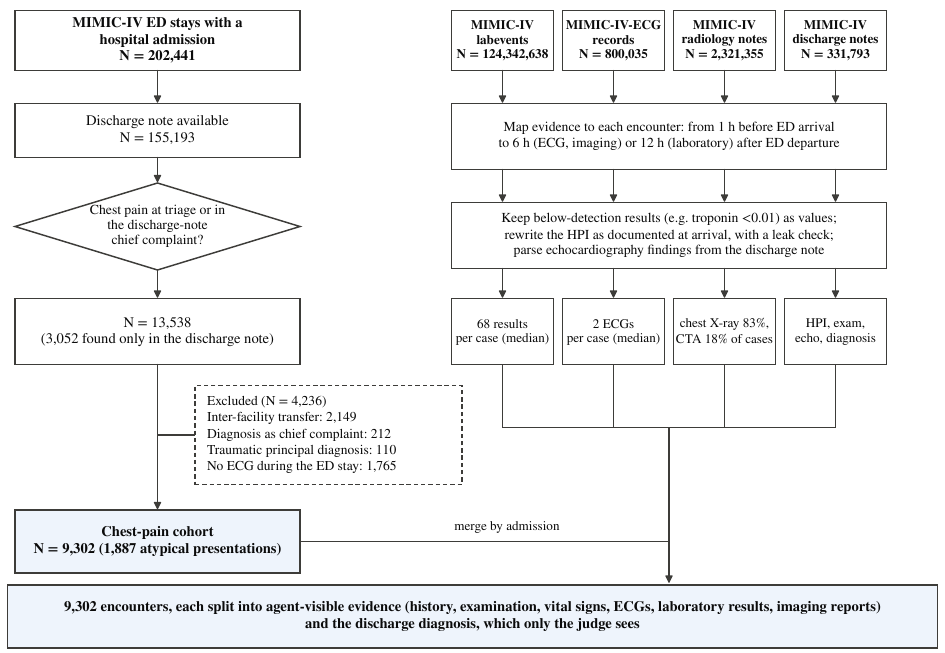}
  \caption{Construction of EDBench-CP. Left: selection of the chest-pain cohort from MIMIC-IV admissions. Right: evidence sources and how each is mapped to an encounter. Counts are hospital admissions.}
  \label{fig:flowchart}
\end{figure*}

\subsection{Cohort Selection} \label{app:cohort}

We draw on four modules of MIMIC-IV~\citep{PhysioNet-mimiciv-3.1}: ED triage and stays, hospital admissions, free-text notes, and diagnostic ECGs. An encounter is included when either the triage chief complaint or the \emph{Chief Complaint} section of the discharge note indicates chest pain, and the admission has a discharge note, which supplies the history, the physical examination and the grading target.

Figure~\ref{fig:flowchart} (left) shows how the 9{,}302 encounters are obtained from MIMIC-IV~\citep{PhysioNet-mimiciv-3.1}. We start from the 202{,}441 ED stays that led to a hospital admission and keep the 155{,}193 whose admission has a discharge note, which supplies the history, the physical examination and the grading target. Of these, 13{,}538 report chest pain in either of two sources: the triage chief complaint, matched case-insensitively against \emph{chest}, the abbreviations \emph{CP} and \emph{C/P}, \emph{substernal}, \emph{precordial} and \emph{CW pain}; or the \emph{Chief Complaint} section of the discharge note, matched against \emph{chest pain}, \emph{chest pressure}, \emph{chest discomfort} and \emph{chest tightness}. For 3{,}052 of them, chest pain appears only in the discharge note. We then exclude 4{,}236 encounters in the four groups described below: 2{,}149 inter-facility transfers, 212 whose chief complaint is itself a diagnosis, 110 with a traumatic principal diagnosis, and 1{,}765 without an ECG during the ED stay. This leaves 9{,}302 encounters, 1{,}887 of which are atypical presentations found only through the discharge note.

Both sources are needed. Selecting on the triage string alone yields only 4{,}546 encounters. It misses abbreviations (``CP'', ``C/P'') and, more importantly, atypical presentations, in which the patient is triaged for dyspnoea, palpitations or weakness and the physician later records chest pain as the chief complaint. These 1{,}887 encounters make up 20.3\% of the final cohort, and 18.8\% of them have a time-critical principal diagnosis, compared with 15.8\% of typical presentations. Their evidence is as complete as that of typical presentations (serial troponin 72.4\% vs.\ 74.1\%, physical examination 97.2\% vs.\ 96.5\%). Because atypical presentation is most common in women, diabetic and elderly patients, dropping these encounters would bias the benchmark towards textbook cases.

We exclude four groups. (1)~Inter-facility transfers, identified by the chief complaint or by the admission-location field, because such patients arrive with a workup already done. The admission-location field catches 547 transfers that the chief-complaint string misses. (2)~Encounters whose chief complaint is itself a diagnosis (``STEMI'', ``NSTEMI'', ``troponin'', ``ACS'', ``angina''). (3)~Encounters with a traumatic principal diagnosis, which follow a separate pathway with little diagnostic uncertainty. (4)~Encounters without an ECG recorded during the ED stay, since a chest-pain workup cannot proceed without one.

The last exclusion does not remove milder cases. The 1{,}765 excluded patients have ICU admission and mortality rates similar to those retained (13.9\% vs.\ 14.5\%, 1.02\% vs.\ 1.30\%), longer stays, and include 160 myocardial infarctions and 43 pulmonary embolisms. For 72\% of those infarctions an ECG exists at another time, so much of the gap reflects linkage of the ECG module rather than clinical practice.

\subsection{Data Collection} \label{app:evidence}

For each encounter we collect everything the treating team obtained: the prior diagnoses, medications and procedures available at arrival; the physical examination; serial vital signs; every ECG with its machine-read report; laboratory results; and radiology reports.

\paragraph{Time windows.} Evidence is collected from one hour before ED arrival, because triage bloods are often charted before the formal arrival time, until six hours after departure for ECGs and imaging and twelve hours after departure for laboratory results. The pre-arrival hour recovers laboratory results for 120 encounters and an ECG for 148. The longer laboratory window captures the repeat troponin required by the 0/3\,h and 0/6\,h rule-out protocols. Coverage of serial troponin rises from 21.5\% with no post-departure window to 61\% at six hours and 74\% at twelve, and only to 77\% at twenty-four, so we stop at twelve.

\paragraph{Below-detection results.} Results such as ``$<$0.01'' are stored in a comment field rather than as numeric values in MIMIC-IV. We parse them, together with above-assay results (``$>$25''), so that a negative troponin is distinguishable from a troponin that was never drawn. With this change 91.2\% of encounters have an interpretable troponin, against 33.1\% when only numeric values are used.

\paragraph{Echocardiography.} MIMIC-IV has no structured echocardiography table. Transthoracic echo findings are taken from the \emph{Pertinent Results} section of the discharge note, restricted to the echo paragraph and cut off before any diagnosis or hospital-course text.

\subsection{At-Arrival History} \label{app:hpi}
\paragraph{Why the history is the starting point.} In the emergency department, the first information a physician acts on is not a test but the patient's account: when the pain started, what it feels like, where it radiates, and what is already known about the patient's heart. This account sets the pretest probability that decides which tests are worth ordering. Chest-pain guidelines place the history first in risk assessment~\citep{gulati2021chestpain}, and it is the first component of the HEART score~\citep{six2008heart}. The triage record alone, a chief complaint of ``chest pain'' with vital signs, cannot support this reasoning: an agent starting from it could do little more than order every test, which would also make the cost of the workup uninformative. We therefore start each episode from the history as it could have been documented at arrival. The only written history in MIMIC-IV is the retrospective one in the discharge summary, so we reconstruct the at-arrival version from it.

The history of present illness in a discharge summary is written after the fact and routinely contains the answer, for example ``started on heparin for possible NSTEMI given ST depressions and troponin 0.08''. We extract the HPI section of each discharge note and rewrite it with an LLM (Qwen 3.8 27B), at temperature 0, as the history a clinician could have documented at the moment of arrival (prompt in Figure~\ref{fig:hpi-prompt}). The rewrite keeps symptom onset, character, radiation, associated symptoms, risk factors and findings from earlier admissions, which the clinician would genuinely know, and removes any result, diagnosis, treatment or ED course belonging to the current episode.

\paragraph{Leakage checks.} The model's own report that the text is clean proved unreliable: it occasionally claimed to have removed all leaked content while leaving the sentence verbatim. A rule-based detector therefore checks every output for test results from the current episode, working diagnoses, treatments and ED-course statements. When it fires, the model is asked again, with the leaked categories named, up to two times.

\begin{table}[t]
\centering
\small
\begin{tabular}{@{}lr@{}}
\toprule
\multicolumn{2}{@{}l}{\textit{Cohort}} \\
Encounters & 9{,}302 \\
\midrule
Atypical presentation & 1{,}887 (20.3\%) \\
Age, median (IQR) & 65 (54--76) \\
Female & 4{,}243 (45.6\%) \\
Triage acuity 1--2 & 7{,}373 (79.3\%) \\
Arrived by ambulance & 4{,}190 (45.0\%) \\
\midrule
\multicolumn{2}{@{}l}{\textit{Evidence available}} \\
At-arrival history & 9{,}275 (99.7\%) \\
Physical examination & 8{,}993 (96.7\%) \\
ECG (ECGs per case, median) & 9{,}302 (2) \\
Troponin, any interpretable result & 8{,}487 (91.2\%) \\
Serial troponin ($\geq$2 draws) & 6{,}857 (73.7\%) \\
Lab results per case, median (IQR) & 68 (52--88) \\
Chest X-ray & 7{,}732 (83.1\%) \\
CT angiography of the chest & 1{,}687 (18.1\%) \\
Echocardiogram & 2{,}420 (26.0\%) \\
\midrule
Full-workup cost, median (IQR)$^\dagger$ & \$408 (228--705) \\
\bottomrule
\end{tabular}
\caption{EDBench-CP cohort statistics. $^\dagger$Cost of ordering every study present in the record once, priced as in Appendix~\ref{app:cost}; estimated on a random sample of 2{,}000 encounters. It is an upper bound that an efficient agent should usually beat.}
\label{tab:stats}
\end{table}

\paragraph{Outcome.} Table~\ref{tab:stats} shows the cohort statistics and the evidence availability of the final cohort.
9{,}275 encounters (99.7\%) receive an at-arrival history; the other 27 have no usable HPI section or failed extraction. Across the cohort the rewrite removed test results in 8{,}356 histories, ED-course statements in 8{,}240, treatments in 7{,}393, diagnoses in 4{,}754 and outcomes in 1{,}349. A further 1{,}888 encounters (20.3\%) are flagged as having an outside workup mentioned in the history, but we think this is reasonable because the hospital should treat the transfer patients with the prior knowledge in real scenario.

\begin{figure*}[tp]
\begin{promptbox}{At-Arrival History Rewrite Prompt}
\pl{Role.} You rewrite clinical histories for a diagnostic benchmark.\\[2pt]
\pl{Context.} You are given the ``History of Present Illness'' from a DISCHARGE SUMMARY. It was written retrospectively, so it often leaks information that was NOT known when the patient first arrived in the emergency department.\\[2pt]
\pl{Instruction.} Rewrite it as the HPI a clinician could have documented AT THE MOMENT OF ED ARRIVAL, before any workup.\\[2pt]
\pl{Keep} (this is the patient's story):
\begin{pitems}
\item age, sex, and relevant past medical/surgical history, medications, risk factors
\item what brought them in: symptom onset, timing, character, location, radiation, severity
\item provoking/relieving factors, associated symptoms, pertinent positives AND negatives
\item what happened before arrival (at home, symptoms in transit, mode of arrival)
\item baseline functional status, allergies, social/family history if mentioned
\end{pitems}
\pl{Remove} (unknowable at arrival); all of this refers to THIS episode:
\begin{pitems}
\item any test result from this presentation: ECG findings, troponin or other lab values, imaging findings, echo
\item any diagnosis, working diagnosis, or diagnostic conclusion for this episode (``ruled in for MI'', ``found to have a PE'', ``c/f ACS'')
\item anything done after arrival: ED vitals, medications given, procedures, consults, admission, transfer for catheterisation, hospital course, outcome
\item workup performed at an outside hospital for THIS episode (you may state that the patient was transferred, but not what was found)
\end{pitems}
\pl{Keep even though it mentions tests} (this is known history, not leakage):
\begin{pitems}
\item findings from a PREVIOUS, already-completed admission or prior outpatient workup (e.g.\ ``during an admission last month he had ST elevations and was treated for pericarditis'', ``known LVEF 35\% on echo in \_\_\_'', ``prior cath showed 90\% LAD''). A clinician genuinely knows these at arrival. Keep them, and make it explicit that they are historical.
\end{pitems}
\pl{Rules.}
\begin{pitems}
\item Do NOT invent anything. Only reuse information present in the source.
\item Keep the de-identification placeholders (\_\_\_) exactly as they appear.
\item Write in the same clinical register, third person, 1 paragraph.
\item If after removal almost nothing remains, output what little is legitimately known.
\end{pitems}
\pl{Input.}
\begin{pitems}
\item History of present illness from the discharge note: \{hpi\}
\end{pitems}
\pl{Output format.} Reply with ONE JSON object and nothing else. \texttt{leak\_types} = what you had to REMOVE; \texttt{outside\_workup} = true if the patient arrived with a diagnosis/workup already established elsewhere.\par\smallskip
{\ttfamily\{"hpi\_at\_arrival": "...",\\
\phantom{\{}"leak\_types": [ ... any of: "test\_results","diagnosis","treatment","ed\_course","outcome" ],\\
\phantom{\{}"outside\_workup": true|false\}}
\end{promptbox}
\caption{Prompt for the at-arrival history rewrite. The system prompt is reproduced with its sections labelled; the input is the HPI section extracted from the discharge note.}
\label{fig:hpi-prompt}
\end{figure*}

\subsection{Label Statistics} \label{app:labels}

The primary label of an encounter is the first diagnosis line of the discharge diagnosis text, after skipping section headers (``Primary diagnosis:'') and separator lines. It can be parsed for 9{,}073 of the 9{,}302 encounters; the rest have an empty diagnosis field or begin directly with secondary diagnoses. Labels are lower-cased and stripped of numbering, bullets and punctuation, and a small synonym table merges abbreviation and spelling variants of the same entity (e.g.\ ``NSTEMI'' and ``non-ST-segment elevation myocardial infarction''; ``pulmonary embolus'' and ``pulmonary embolism''). Clinically distinct labels, such as ``chest pain'' and ``atypical chest pain'', are left apart. Multi-line primary diagnoses contribute only their first line. A label counts as time-critical in Figure~\ref{fig:labels} when it names one of the diagnoses in \S\ref{sec:discussion} (keyword match, including variants such as ``pulmonary embolus''), unless the label negates or hedges it (``chest pain without myocardial infarction'', ``possible pulmonary embolism''). Table~\ref{tab:labels} lists the 30 most frequent labels.

Note that a primary diagnosis of "chest pain" is a legitimate discharge diagnosis: it records that the workup excluded the dangerous causes without identifying a specific one, and 70\% of these encounters carry a symptom code (ICD-10 R07) as their principal billing code. Because only the first line of the primary diagnosis is used, a few encounters whose note lists the symptom before its cause are also counted here.
\begin{table*}[h]
\centering
\small
\begin{tabular}{@{}rlrr@{\hspace{2.2em}}rlrr@{}}
\toprule
\# & Primary diagnosis & $n$ & \% & \# & Primary diagnosis & $n$ & \% \\
\midrule
1 & Chest pain & 673 & 7.4 & 16 & Stable angina & 45 & 0.5 \\
2 & NSTEMI$^\dagger$ & 469 & 5.2 & 17 & Acute on chronic systolic heart failure & 45 & 0.5 \\
3 & Atypical chest pain & 353 & 3.9 & 18 & Hypertensive emergency & 44 & 0.5 \\
4 & Coronary artery disease & 329 & 3.6 & 19 & Atrial fibrillation with RVR & 44 & 0.5 \\
5 & Unstable angina & 281 & 3.1 & 20 & Hypertensive urgency & 41 & 0.5 \\
6 & Pulmonary embolism$^\dagger$ & 179 & 2.0 & 21 & Costochondritis & 40 & 0.4 \\
7 & STEMI$^\dagger$ & 129 & 1.4 & 22 & Musculoskeletal chest pain & 40 & 0.4 \\
8 & Non-cardiac chest pain & 107 & 1.2 & 23 & Urinary tract infection & 38 & 0.4 \\
9 & Pneumonia & 102 & 1.1 & 24 & Abdominal pain & 38 & 0.4 \\
10 & Community-acquired pneumonia & 99 & 1.1 & 25 & Asthma exacerbation & 35 & 0.4 \\
11 & COPD exacerbation & 91 & 1.0 & 26 & Pericardial effusion & 29 & 0.3 \\
12 & Pericarditis & 82 & 0.9 & 27 & Syncope & 29 & 0.3 \\
13 & Atrial fibrillation & 68 & 0.7 & 28 & Alcohol withdrawal & 29 & 0.3 \\
14 & Angina & 58 & 0.6 & 29 & Acute cholecystitis & 29 & 0.3 \\
15 & Acute on chronic diastolic heart failure & 51 & 0.6 & 30 & Myopericarditis & 26 & 0.3 \\
\bottomrule
\end{tabular}
\caption{The 30 most frequent primary discharge diagnoses among the 9{,}073 encounters whose primary diagnosis could be parsed, after normalisation and merging of abbreviation variants. Together they cover 39.9\% of encounters; the remaining 60.1\% are spread over 3{,}359 further labels. $^\dagger$Time-critical (can't-miss) diagnosis.}
\label{tab:labels}
\end{table*}

\subsection{Can't Miss Diseases Statistics}

Table~\ref{tab:timecritical} breaks down the encounters with a time-critical principal diagnosis, which form the denominator of the dangerous-miss rate. The set follows the life-threatening causes of acute chest pain that guidelines single out for the emergency physician to identify or exclude~\citep{gulati2021chestpain}: acute myocardial infarction, pulmonary embolism, aortic dissection, pneumothorax and cardiac tamponade. These diagnoses share three properties that make a miss qualitatively worse than other errors. Each carries a high short-term mortality if untreated; each has a time-dependent treatment, namely reperfusion, anticoagulation or thrombolysis, emergency surgery, decompression or drainage; and each can be confirmed or excluded in the ED with a test in the action space (ECG and serial troponin, CT angiography, chest X-ray, echocardiography), so a miss reflects the workup rather than the limits of the record. Unstable angina is not included: it is managed urgently, but without myocardial necrosis its immediate risk is lower, and its diagnosis rests on the history rather than on a confirmatory test. Oesophageal perforation, often listed alongside these causes, is left out because it is the principal diagnosis of only four encounters.

\subsection{Data Example} \label{app:example}

Figure~\ref{fig:example} shows one encounter from the evaluation subset, abridged: the presentation the agent sees at arrival, the evidence released by one possible sequence of actions, and the reference diagnosis that only the judge sees. Measurements taken more than once, such as vital signs and blood draws, are released one at a time, so following a trend takes several turns. MIMIC-IV replaces identifiers with ``\_\_\_''. The encounter illustrates why the workup cannot stop once acute coronary syndrome is excluded: both troponins are negative, while the D-dimer, the CT angiogram and the echocardiogram point to a pulmonary embolism.

\begin{figure*}[tp]
\begin{databox}{Example Encounter}
\pl{Presentation at arrival} (visible at the start of the episode)
\begin{pitems}
\item 55-year-old man; chief complaint ``cp''; triage acuity 2; pain 10/10; arrived walking.
\item Triage vital signs: temperature 98.4\,\textdegree F, heart rate 124, respiratory rate 20, SpO$_2$ 98\%, blood pressure 166/98.
\item History of present illness (rewritten as at arrival): ``\_\_\_ c/o exertional, pleuritic substernal chest pain x 3 wks. The pain comes on after walking \textasciitilde 1 block, resolves with rest after 5 minutes. Associated with SOB, no diaphoresis, N/V. Denies cough, F/C. No prior DVT/PE. No recent travel or immobility or trauma. Also c/o of chronic L sided neck and lower back pain. He saw his PCP for these symptoms and was advised to go to the ED.''
\end{pitems}
\tcbline
\pl{Evidence released on request}: one possible sequence of actions; each request releases only the next unseen result, so repeated measurements need repeated requests.
\begin{enumerate}[leftmargin=1.7em, itemsep=0pt, parsep=0pt, topsep=1pt, label=\arabic*.]
\item \texttt{ecg} (\$15): Sinus tachycardia. Indeterminate axis. Inferior infarct, age undetermined. Anterior T wave changes are nonspecific. Abnormal ECG.
\item \texttt{labs troponin} (\$45), blood draw 1 of 4: troponin T $<$0.01\,ng/mL; D-dimer 2{,}062\,ng/mL (reference 0--500); creatinine 0.7\,mg/dL; hemoglobin 15.0\,g/dL; white blood cells 7.8\,K/\textmu L; \ldots\ (36 results)
\item \texttt{vitals} (\$0), recording 1 of 3: heart rate 104, respiratory rate 18, SpO$_2$ 94\%, blood pressure 145/87.
\item \texttt{imaging cxr} (\$33): ``The lungs are well expanded and clear. There is no pleural effusion or pneumothorax. \ldots\ IMPRESSION: No acute intrathoracic process.''
\item \texttt{history} (\$0): prior diagnoses: anxiety state; backache; chest pain, unspecified; depressive disorder; other chronic pain; \ldots\ Home medications: acetaminophen; gabapentin; hydromorphone; \ldots
\item \texttt{exam} (\$0): General: well appearing male in NAD. Lungs: clear bilaterally, respirations unlabored. Heart: RRR, no murmurs, no loud P2. Extremities: warm, no edema. \ldots
\item \texttt{vitals} (\$0), recording 2 of 3: heart rate 94, respiratory rate 18, SpO$_2$ 96\%, blood pressure 126/84.
\item \texttt{labs coags} (\$45), blood draw 3 of 4: INR 1.0; PT 11.2\,s; PTT 28.6\,s.
\item \texttt{imaging cta} (\$281): ``IMPRESSION: 1. Pulmonary emboli involving all lobes with suggestion of right heart strain which can be further assessed with echocardiogram. 2. Scattered sub 4 mm pulmonary nodules \ldots''
\item \texttt{labs troponin} (\$45), next draw with troponin, 4 of 4: troponin T $<$0.01\,ng/mL; NT-proBNP 796\,pg/mL (reference 0--177); creatinine 0.7\,mg/dL; \ldots
\item \texttt{vitals} (\$0), recording 3 of 3: heart rate 93, respiratory rate 16, SpO$_2$ 92\%, blood pressure 145/96.
\item \texttt{imaging echo} (\$197): ``IMPRESSION: Right ventricular cavity enlargement with free wall hypokinesis. Pulmonary artery hypertension. \ldots''
\item \texttt{diagnose}: ends the episode with the agent's primary and secondary diagnoses.
\end{enumerate}
\tcbline
\pl{Reference} (seen only by the judge)
\begin{pitems}
\item Discharge diagnosis: ``Pulmonary embolism, unprovoked''. Time-critical: yes.
\end{pitems}
\end{databox}
\caption{One encounter of EDBench-CP. The agent starts from the presentation and must request each piece of evidence; the reference diagnosis is never shown to it.}
\label{fig:example}
\end{figure*}

\begin{table}[h]
\centering
\small
\begin{tabular}{@{}lrr@{}}
\toprule
Principal diagnosis & Cohort & Subset \\
\midrule
NSTEMI & 831 (8.9\%) & 58 \\
STEMI & 122 (1.3\%) & 14 \\
Acute MI, site-coded or type 2$^\ast$ & 182 (2.0\%) & 16 \\
Pulmonary embolism & 290 (3.1\%) & 40 \\
Pneumothorax & 56 (0.6\%) & 4 \\
Aortic dissection & 26 (0.3\%) & 7 \\
\midrule
Total & 1{,}507 (16.2\%) & 139 \\
\bottomrule
\end{tabular}
\caption{Time-critical (can't-miss) principal diagnoses in the cohort ($n=9{,}302$) and in the 400-encounter evaluation subset; these form the denominator of the dangerous-miss rate. $^\ast$Acute MI coded by site in ICD-9 (STEMI and NSTEMI are not separated) or as type~2 MI. No encounter has cardiac tamponade as its principal diagnosis.}
\label{tab:timecritical}
\end{table}

\section{LLM-as-a-Judge} \label{app:judge}

\paragraph{Why an LLM judge.} Grading in EDBench-CP compares two pieces of free text, the agent's diagnosis list and the physician's discharge diagnosis, and four properties of this comparison rule out simpler metrics.
\begin{itemize}[leftmargin=1.2em, itemsep=1pt, topsep=2pt]
\item \emph{No closed label set.} Even after normalisation the primary discharge diagnoses span 3{,}389 distinct labels (Appendix~\ref{app:labels}), so exact match or classification accuracy against a fixed list is not available.
\item \emph{Equivalence is clinical, not lexical.} ``STEMI'' and ``acute ST-elevation myocardial infarction'' name the same entity, and a more or less specific term may be equally acceptable; conversely, a type~A intramural haematoma and a descending aortic dissection share most of their words but lead to surgery in one case and medical management in the other.
\item \emph{The reference lists the whole admission.} A discharge summary records every problem managed in hospital, so the grader must first find the entry that explains the chest pain and must not reward a list for matching unrelated comorbidities.
\item \emph{Scale.} The reported results alone contain 15{,}600 graded episodes (39 runs of 400 encounters), far beyond what physicians could score by hand.
\end{itemize}

\begin{table}[h]
\centering
\small
\setlength{\tabcolsep}{4pt}
\begin{tabular}{@{}lcccc@{}}
\toprule
 & \multicolumn{2}{c}{Pearson $r$} & \multicolumn{2}{c}{Danger AUC} \\
\cmidrule(lr){2-3} \cmidrule(l){4-5}
Metric & Full & Primary & Full & Primary \\
\midrule
BLEU-4 & 0.11 & -- & 0.65 & -- \\
TF-IDF cosine & 0.27 & -- & 0.62 & -- \\
Token-level F1 & 0.32 & 0.52 & 0.70 & 0.79 \\
ROUGE-1 recall & 0.40 & 0.53 & 0.73 & 0.78 \\
Category match & 0.36 & 0.41 & 0.66 & 0.66 \\
\midrule
LLM judge (ours) & \textbf{0.74} & -- & \textbf{0.91} & -- \\
\bottomrule
\end{tabular}
\caption{Agreement of automatic metrics with a physician's blind scores on 400 diagnoses produced by one agent. Pearson $r$ is computed against the 1--5 physician score, on the \textbf{full diagnosis list} or on the \textbf{primary diagnosis} only. Danger AUC measures how well a low metric value flags the time-critical encounters that the physician scored $\leq 2$ ($n=139$). Category match compares keyword-assigned diagnostic categories. The judge always reads the full list.}
\label{tab:metrics}
\end{table}

Lexical metrics fail on exactly these points, which we measure against physician judgement (Table~\ref{tab:metrics}). We took the 400 encounters of the evaluation subset and the diagnosis list that one agent submitted for each, produced in an earlier version of the environment. A physician scored every list against the discharge diagnosis on the same five-point scale. Each pair was shown in the English original alongside a machine translation into the physician's working language, and the judge's score and justification were withheld, so the physician scored blind. We then scored the same 400 pairs with each candidate metric. The lexical metrics share one tokenisation: lower-cased alphanumeric tokens of at least three characters, with stop words and generic clinical modifiers such as ``acute'', ``chronic'' and ``unspecified'' removed. \textbf{ROUGE-1 recall} is the share of reference tokens that appear in the submission; \textbf{token-level F1} is the harmonic mean of token precision and recall; \textbf{BLEU-4} is the geometric mean of 1- to 4-gram precisions with a brevity penalty; \textbf{TF-IDF cosine} uses document frequencies over the 800 texts; and category match is 1 when both lists fall into the same diagnostic category under the keyword rules used for the cohort statistics. Where it is meaningful, each metric is computed both on the full lists and on the primary diagnosis alone. \textbf{Pearson $r$} is taken between the metric value and the physician score over all 400 encounters. The danger AUC is restricted to the 139 time-critical encounters, treats a physician score $\leq 2$ as a dangerous miss, and is the probability that a randomly chosen miss receives a lower metric value than a randomly chosen non-miss, with ties counting one half. The LLM-judge row grades the same 400 lists with the final rubric. The best lexical metric, ROUGE-1 recall over the primary diagnosis, reaches $r=0.53$ and an AUC of 0.78, against $r=0.74$ and 0.91 for the judge. The failures run in both directions: an abbreviation scores zero overlap with its expansion although the physician gives full marks, while a list that misses a pulmonary embolism but repeats the patient's comorbidities reaches a ROUGE-1 recall of 0.75 against a physician score of 1.

\paragraph{Grading procedure.} Each episode is graded once, independently of every other episode, in five steps.
\begin{enumerate}[leftmargin=1.4em, itemsep=1pt, topsep=2pt]
\item \emph{Input.} The judge receives the discharge diagnosis list as the target, with the ED principal diagnosis and the admission principal billing code as context that need not be matched, followed by the submitted list. Every submission is presented as an ED physician's list; the judge sees neither the transcript nor the cost, nor whether a model or a clinician wrote the diagnosis.
\item \emph{Locate the target.} Following the rubric (Table~\ref{tab:rubric}, Figure~\ref{fig:judge-prompt}), the judge identifies which reference entry is the cause of the chest pain and grades against that entry only.
\item \emph{Score.} The submitted primary diagnosis sets the band: the correct cause as primary scores 4 or 5, depending on the coverage of secondary diagnoses; the cause listed only as secondary, or a sub-type confusion, scores 3; a wrong but clinically adjacent primary scores 2; and a primary that changes management, or any omission of a time-critical diagnosis present in the reference, scores 1.
\item \emph{Parse.} The judge writes a one-sentence justification and a final line \texttt{SCORE: <1-5>}. If no score can be parsed, which happens when a reasoning model exhausts its output budget on hidden reasoning, the episode is graded once more with twice the budget. Every episode in the reported results received a score.
\item \emph{Aggregate.} The episode score $J$ enters the metrics of Accuracy: an episode is correct if $J \geq 4$, and a time-critical encounter with $J \leq 2$ counts as a dangerous miss.
\end{enumerate}
The rubric was revised once against physician scores on 400 diagnoses so that the judge's score distribution matches the physician's; the same judge and rubric grade every run in Table~\ref{tab:main} except the one marked there.

\begin{table*}[t]
\centering
\small
\begin{tabular}{@{}c >{\raggedright\arraybackslash}p{4.7cm} >{\raggedright\arraybackslash}p{4.4cm} >{\raggedright\arraybackslash}p{4.9cm}@{}}
\toprule
Score & Criterion & Reference (discharge diagnosis) & Submitted list \\
\midrule
5 & Chest-pain cause named as primary; relevant secondary diagnoses covered. & Primary: ST-elevation MI; \emph{E.\ coli} bacteremia. Secondary: coronary artery disease. & Primary: acute anterior ST-elevation MI. Secondary: severe three-vessel coronary artery disease; \ldots \\
\addlinespace
4 & Chest-pain cause named as primary; secondary coverage thin, or extra differentials listed. & Biomarker-negative unstable angina; coronary artery disease with in-stent restenosis; hypertension; \ldots & Primary: unstable angina. Secondary: paroxysmal atrial fibrillation; hypertension; \ldots\ (restenosis omitted) \\
\addlinespace
3 & Correct cause listed only as secondary, or a sub-type confusion of the right diagnosis (e.g.\ NSTEMI vs.\ unstable angina). & Primary: non-ST-elevation MI. Secondary: asthma; hypertension. & Primary: acute decompensated heart failure. Secondary: coronary artery disease (s/p inferior NSTEMI, stents); \ldots \\
\addlinespace
2 & Primary wrong but clinically adjacent; ED management broadly similar. & Costochondritis; atrial fibrillation; acute complicated cystitis; \ldots & Primary: undifferentiated non-cardiac chest pain with hypotension. Secondary: new atrial fibrillation; \ldots \\
\addlinespace
1 & Primary wrong with a different management path; or a time-critical diagnosis in the reference not named anywhere; or no diagnosis. & ST-elevation myocardial infarction. & Primary: acute myocarditis. Secondary: thrombocytopenia; mild left ventricular systolic dysfunction. \\
\bottomrule
\end{tabular}
\caption{Judge rubric with one example per score. At score 5 the co-primary \emph{E.\ coli} bacteremia is not required because it is not the cause of the chest pain. $J \geq 4$ counts as correct.}
\label{tab:rubric}
\end{table*}

\begin{figure*}[tp]
\begin{promptbox}{Judge Prompt}
\pl{Role.} You grade an ED physician's FINAL DIAGNOSIS LIST for a chest-pain encounter against the ground-truth discharge diagnosis, on a 1--5 scale.\\[2pt]
\pl{What you are grading.} Did the physician identify THE CAUSE OF THIS CHEST-PAIN PRESENTATION? That is the question the ED had to answer.\\[2pt]
\pl{Reading the ground truth.} MIMIC discharge summaries often list SEVERAL co-primary diagnoses, because they record every problem managed during the whole admission (e.g.\ ``STEMI'' AND ``E.\ coli bacteremia''). Identify which ground-truth entry is the CHEST-PAIN CAUSE and grade against THAT. Do NOT lower the score because the physician omitted other in-hospital problems that were not the presenting complaint; those are bonus, not requirements.\\[2pt]
\pl{Rules.}
\begin{pitems}
\item Accept wording differences and equivalents (``PE'' == ``pulmonary embolism'') and a reasonably more/less specific term for the same entity.
\item ``Undifferentiated / non-cardiac chest pain'' is a correct PRIMARY when the discharge primary is itself a non-specific chest-pain diagnosis (no acute cause found).
\item The ``admission billing code'' is context only; do NOT require matching it.
\item Listing unrelated comorbidities does NOT lift a 1 to a 2. The question is whether the cause of THIS chest-pain presentation was found, not how long the list is.
\end{pitems}
\pl{Scale} (anchor on THE CHEST-PAIN CAUSE):
\begin{pitems}
\item[\textbf{5}] correctly identifies the chest-pain cause as Primary; relevant secondary diagnoses reasonably covered. Unrelated chronic comorbidities that the ED would not work up (depression, osteoporosis, hypothyroidism\ldots) are NOT required.
\item[\textbf{4}] correctly identifies the chest-pain cause as Primary; secondary coverage thin, or extra differentials listed. Extra diagnoses cost a point only when they are clearly wrong or would lead to harmful management.
\item[\textbf{3}] the correct chest-pain cause APPEARS IN THE LIST but is demoted to Secondary, OR the named entity is a genuine sub-type/variant confusion of the right diagnosis (e.g.\ NSTEMI vs unstable angina).
\item[\textbf{2}] Primary is wrong but CLINICALLY ADJACENT: same disease family and the immediate ED management would be broadly similar (e.g.\ truth unstable angina, answer stable angina; truth pneumonia, answer bronchitis).
\item[\textbf{1}] Primary is wrong and would send the patient down a different management path, OR any can't-miss diagnosis (STEMI/NSTEMI, pulmonary embolism, aortic dissection, pneumothorax, tamponade) present in the ground truth is not named anywhere in the list, OR no diagnosis was given.
\end{pitems}
\pl{Input.}
\begin{pitems}
\item Ground truth: discharge diagnosis list (target); ED principal diagnosis; admission principal billing code (context only)
\item Physician's final diagnosis list: \{final\_dx\}
\end{pitems}
\pl{Output format.} A one-sentence justification, then a final line exactly:
{\ttfamily SCORE: <1-5>}
\end{promptbox}
\caption{Prompt of the judge (rubric in Table~\ref{tab:rubric}). The system prompt is reproduced with its sections labelled; the input is the user turn.}
\label{fig:judge-prompt}
\end{figure*}

\begin{figure}[h]
  \centering
  \includegraphics[width=0.7\textwidth]{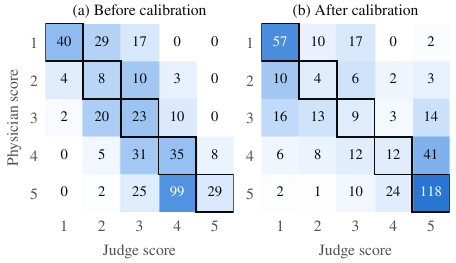}
  \caption{Judge versus physician scores on 400 diagnoses, before and after calibration of the rubric. }
  \label{fig:confusion}
\end{figure}

\paragraph{Calibration against a physician.} We compared the results of the Qwen 3.8 27B model with the ground truth and had them scored by the Claude Sonnet 5.5 model and an emergency department physician with 12 years of experience, respectively. Based on the differences in scoring, we modified our rubrics to make the LLM’s scoring more closely align with that of human physicians. Figure~\ref{fig:confusion} shows both confusion matrices and Table~\ref{tab:calibration} the agreement statistics.

Before calibration the judge was systematically stricter than the physician, by 0.32 points on average. It compressed the scale towards the middle and rarely awarded full marks: of the 155 diagnoses the physician scored 5, it gave 99 a 4. Its errors were therefore small and consistent, mostly one point below the physician, which yields high agreement within one point but exact agreement on only a third of the diagnoses.

After calibration the mean bias falls to $-0.03$ and exact agreement rises to one half. The two ends of the scale now agree closely: the judge gives 5 to 118 of the 155 diagnoses the physician scored 5, and 1 to 57 of the 86 the physician scored 1. The remaining disagreement sits in the middle. Of the 55 diagnoses the physician scored 3, the judge agrees on only 9 and moves the rest towards the ends, 29 down to 1 or 2 and 17 up to 4 or 5, and it scores 41 of the physician's 79 fours as 5. The calibrated judge is thus more decisive than the physician: it is right more often but, when wrong, wrong by more, so agreement within one point drops slightly while quadratic-weighted $\kappa$ improves. Partially correct diagnoses are where the judge is least reliable.

These differences matter for the reported metrics in different ways. Accuracy ($J \geq 4$) moves by up to 12.5 points across the three graders, because it depends on where the boundary between 3 and 4 falls. The dangerous-miss rate moves by only 2.2 points, because a clearly wrong answer to a time-critical encounter is recognised by every grader. This is why we treat the dangerous-miss rate as the more robust comparison across studies.

\begin{table}[h]
\centering
\small
\begin{tabular}{@{}lccc@{}}
\toprule
 & Physician & Before & After \\
\midrule
Mean score & 3.48 & 3.16 & 3.45 \\
Bias vs.\ physician & -- & $-0.32$ & $-0.03$ \\
Exact agreement & -- & 33.8\% & 50.0\% \\
Within one point & -- & 86.5\% & 79.8\% \\
Off by two or more & -- & 13.5\% & 20.3\% \\
Pearson $r$ & -- & 0.77 & 0.74 \\
Quadratic-weighted $\kappa$ & -- & 0.72 & 0.74 \\
\midrule
Accuracy ($J \geq 4$) & 58.5\% & 46.0\% & 54.8\% \\
Dangerous-miss rate & 29.5\% & 28.1\% & 27.3\% \\
\bottomrule
\end{tabular}
\caption{Agreement between the judge and a physician on 400 diagnoses before and after calibration, and the resulting benchmark metrics. }
\label{tab:calibration}
\end{table}

\section{Cost Model} \label{app:cost}

Prices follow the 2025/26 Medicare national average: the non-facility Physician Fee Schedule~\citep{cms2026pfs} for ECG (\$15) and imaging (chest X-ray \$33 to CT angiography \$281), and the Clinical Laboratory Fee Schedule~\citep{cms2026clfs} for laboratory tests. Because a laboratory order returns the whole blood draw, pricing each panel separately would only reward guessing the cheapest panel name for an identical result. We therefore charge a flat \$45 per blood draw, the sum of the fee-schedule prices for a standard initial ED draw (troponin, basic metabolic panel, complete blood count, coagulation); urinalysis is a separate specimen at \$5. History, examination and vital signs are free, because they are covered by the ED visit charge. Medicare prices are public and identical across sites, which keeps costs comparable between studies. Hospital charges are typically three to four times higher, but a uniform scaling does not change any comparison.

\FloatBarrier
\section{Experimental Environment} \label{app:setup}

All evaluated models are served with vLLM~\citep{kwon2023vllm} (version 0.29) through an OpenAI-compatible endpoint on a local server with four NVIDIA H200 GPUs. Every model is served with a maximum context length of 100{,}000 tokens; shorter contexts truncated long episodes in pilot runs.

\paragraph{Episode control.} An episode ends when the agent submits a diagnosis or after $T=20$ steps; at the last step the agent is instructed to diagnose. A reply that contains neither an action nor a diagnosis is answered with a format reminder and does not count as a step; after three such replies the episode ends without a diagnosis, which the judge scores as 1.

\paragraph{ECG images.} For multimodal models, each released ECG is accompanied by the 12-lead tracing rendered from the MIMIC-IV-ECG waveform at 25\,mm/s and 10\,mm/mV on a standard grid. All other observations are identical to the text-only setting.

\paragraph{Anchors.} The blind and oracle anchors are single-turn: the model receives the presentation alone (blind) or the presentation together with every evidence item of the encounter (oracle), and replies with at most two sentences of reasoning followed by its diagnosis list.

\FloatBarrier
\section{Additional Results} \label{app:results}

\begin{figure*}[t]
  \centering
  \includegraphics[width=\textwidth]{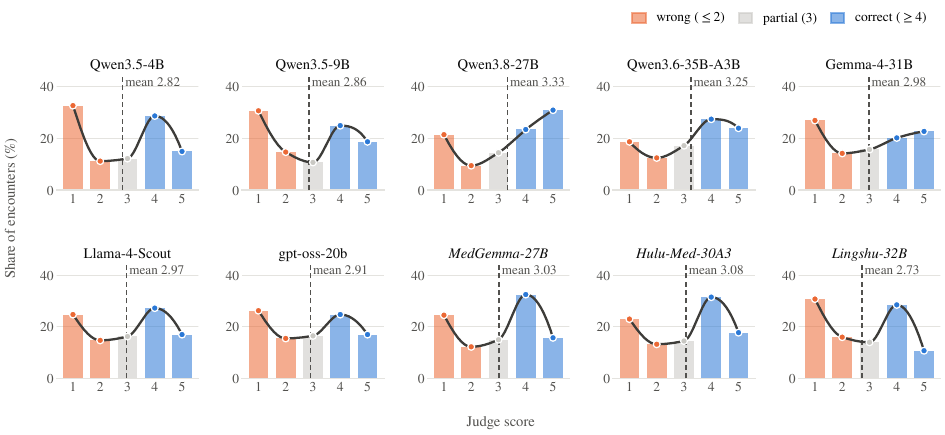}
  \caption{Distribution of judge scores per model in the gated setting with text-only input (400 encounters each). Medical models in italics.}
  \label{fig:dist-text}
\end{figure*}

\begin{figure*}[t]
  \centering
  \includegraphics[width=\textwidth]{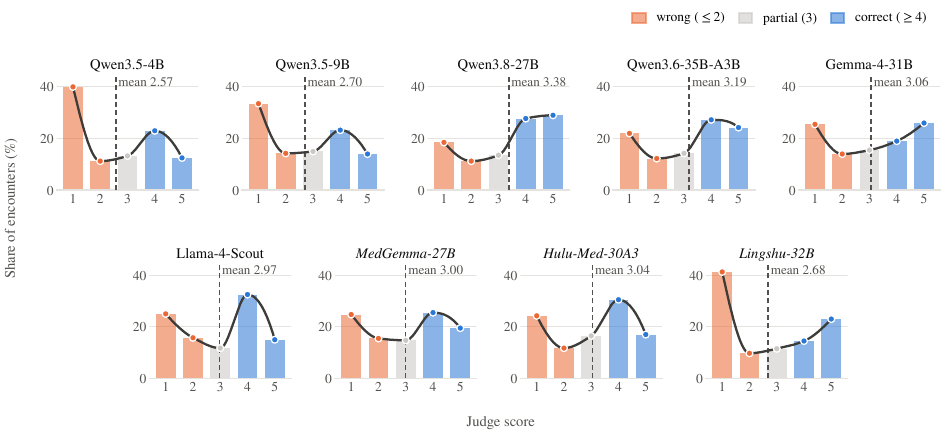}
  \caption{Distribution of judge scores per multimodal model in the gated setting with ECG images. Medical models in italics.}
  \label{fig:dist-vlm}
\end{figure*}

\paragraph{Score distributions.} Figures~\ref{fig:dist-text} and~\ref{fig:dist-vlm} show how the judge scores of each model are distributed in the gated setting, with text-only input and with ECG images.

Every distribution is bimodal, with one peak at score 1 and another at 4 or 5, and a thinner middle. Agents rarely land on a partially correct answer: they either name the cause of the chest pain or miss it, and the calibrated judge, which is more decisive than the physician in the middle of the scale (Appendix~\ref{app:judge}), sharpens this further. Models differ mainly at the two ends. The weakest models fail outright on about a third of the encounters (a score of 1 for 31--33\% of Qwen3.5-4B, Qwen3.5-9B and Lingshu-32B), against about a fifth for Qwen3.6-35B-A3B and Qwen3.8-27B. Among correct answers, most models give a 4 more often than a 5, naming the cause as the primary diagnosis but covering the secondary diagnoses thinly. The gap is largest for the medical models, where MedGemma-27B and Lingshu-32B give two to two and a half times as many 4s as 5s. With text-only input, Qwen3.8-27B is the only model whose most frequent score is 5, which is what lifts its mean score above that of Qwen3.6-35B-A3B at similar accuracy.

We would like to explain here why we use ECG images instead of other images. We had discussions with physicians, who believed that the information contained in image reports is sufficient for analysis, whereas ECG, because it is extracted directly from the ECG machine, lacks key diagnoses and is clearly misleading. Moreover, in the chest pain cohort, ECG images are crucial in the decision-making process and serve as the diagnostic basis for many critical conditions. With ECG images, the distributions of most models barely change. The exceptions are Qwen3.5-4B, which fails outright on 40\% of encounters instead of 33\%, and Lingshu-32B, which gives more scores of both 1 and 5. It can be seen that the VLM model with images added did not change much; this may be because the model's visual token effect is poor, or because the model has not been trained enough on ECGs and therefore cannot provide additional information.

\begin{figure*}[t]
  \centering
  \includegraphics[width=0.72\textwidth]{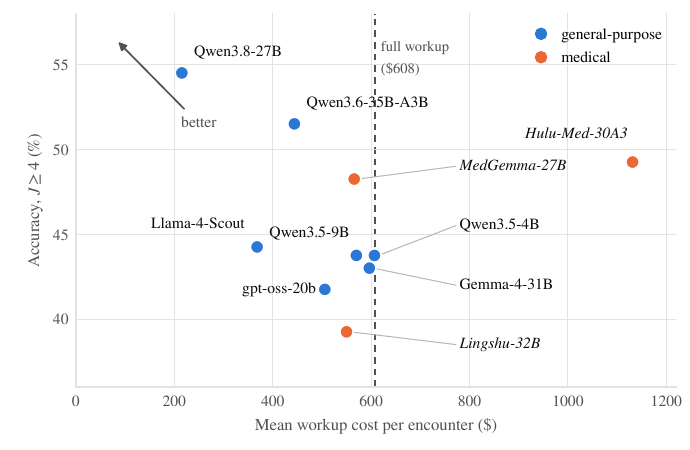}
  \caption{Accuracy against mean workup cost per encounter for each agent in the gated setting with text-only input. The dashed line marks the mean cost of ordering every study in the record once. Medical models in italics.}
  \label{fig:cost-accuracy}
\end{figure*}

\begin{figure*}[t]
  \centering
  \includegraphics[width=\textwidth]{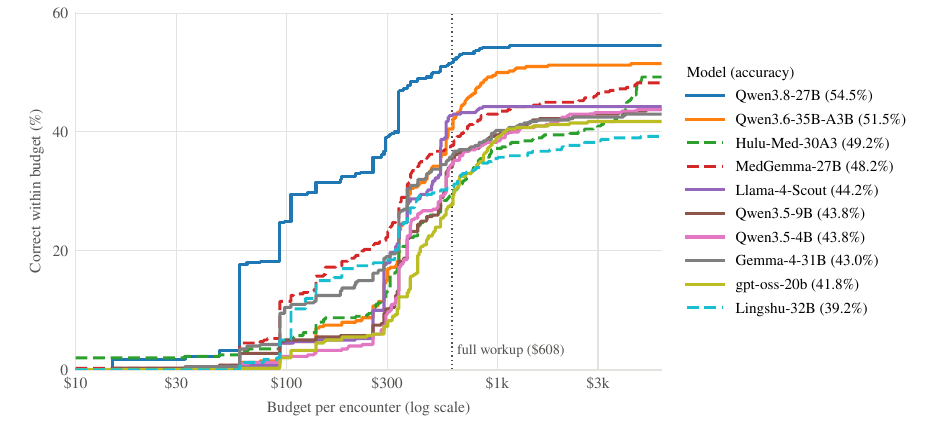}
  \caption{Share of the 400 encounters diagnosed correctly ($J \geq 4$) at no more than a given cost per encounter, gated setting with text-only input. Each curve ends at the model's accuracy, given in the legend; medical models are dashed, and the dotted line marks the mean cost of the full workup.}
  \label{fig:budget}
\end{figure*}

\paragraph{Cost and accuracy.} Figure~\ref{fig:cost-accuracy} plots each agent's accuracy against its mean workup cost with text-only input. The full workup, ordering every study in the record once, costs \$608 per encounter on average. Six of the ten agents spend between \$500 and \$610, close to the full workup, and none of them reaches 50\% accuracy; their budget curves stay low until about \$200 and then rise steeply up to the cost of the full workup, the pattern of a fixed battery of blood tests and imaging ordered for most patients. Hulu-Med-30A3 spends \$1{,}131, almost twice the full workup, and reaches a fifth of its correct diagnoses only beyond \$1{,}600, because repeated orders for results that do not exist are charged every time (\S\ref{sec:discussion}). Figure~\ref{fig:budget} stratifies the same runs by cost, showing for each budget the share of all encounters that the agent diagnosed correctly at no more than that cost.
Qwen3.8-27B is the exception: it is both the most accurate and the cheapest agent, at about a third of the full workup. With \$100 per encounter it has already diagnosed a quarter of the encounters correctly, while no other model exceeds 12\%, and against the blind and oracle anchors it recovers 74\% of the gain that full evidence offers, compared with 64\% for Qwen3.6-35B-A3B at \$444 and 27\% for Gemma-4-31B, which spends 98\% of the full workup. Spending more does not buy accuracy; what separates the agents is which tests they choose. Figure~\ref{fig:budget} stratifies the same runs by cost, showing for each budget the share of all encounters that the agent diagnosed correctly at no more than that cost.

\begin{figure*}[tp]
\centering
\includegraphics[width=\textwidth]{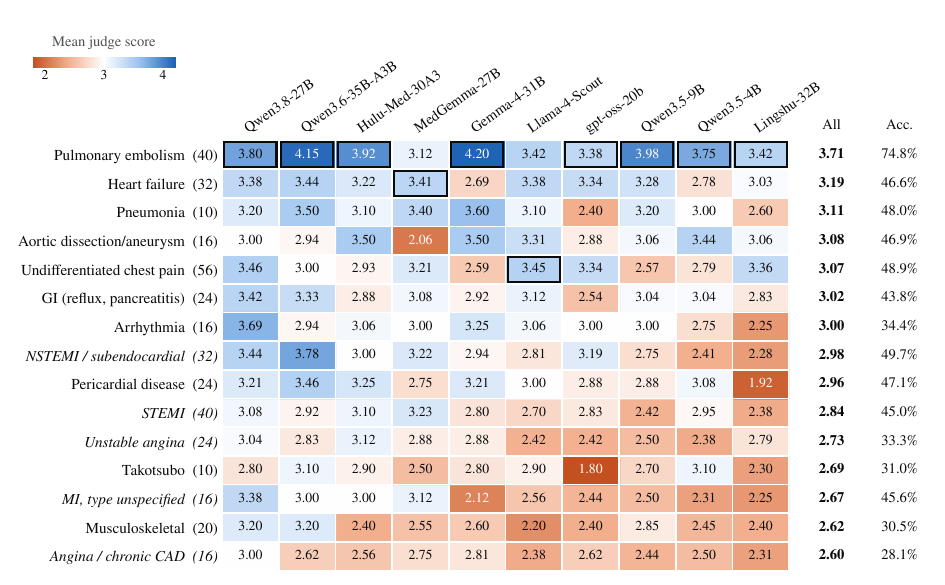}
\caption{Mean judge score of each agent by category of the reference diagnosis, gated setting with text-only input. The number of encounters is given after each category; categories with fewer than ten encounters are omitted (24 of 400). The outlined cell in each column is that agent's best category. \emph{All} and \emph{Acc.} pool the ten agents. Coronary categories and medical models are set in italics; agents are ordered by overall mean score.}
\label{fig:bycategory}
\end{figure*}

\paragraph{Results by diagnostic category.} Figure~\ref{fig:bycategory} breaks the gated text-only runs down by the category of the reference diagnosis. Two patterns hold across agents; beyond them, agents differ in which categories they find hard, but less than the per-agent figures suggest.

Pulmonary embolism is the easiest category by a wide margin: 3.71 against 3.19 for the next, and 74.8\% accuracy where no other category reaches 50\%. It ranks first for eight of the ten agents and within the top three for nine. Its workup ends in a single decisive study, and agents order it: a CT angiogram appears in 77\% of these episodes, which are correct in 84\% of cases, against 44\% when no angiogram is ordered.

Acute coronary syndrome and its variants behave in the opposite way. Pooling STEMI, NSTEMI, unstable angina, chronic coronary disease and myocardial infarction of unspecified type (128 of 400 encounters) gives 2.80 and 42.0\% accuracy, against 3.09 and 47.8\% for every other category combined, and the family scores below the agent's own overall mean for every agent except MedGemma-27B. The gap does not come from partial credit for naming the family without the sub-type: a score of 3 is as common as elsewhere (14.2\% against 14.9\%). It comes from outright misses, scored 1 in 33.7\% of coronary gradings against 22.4\% for the rest, and in 45.0\% for myocardial infarction of unspecified type, consistent with the troponin failures described in main text. STEMI (40 encounters), the category an ECG is most expected to settle, reaches only 2.84. Giving the nine multimodal agents the ECG trace changes the outcome of 82 of their 360 STEMI episodes, in both directions, and leaves accuracy at 45.0\%; these misses are not resolved by letting the agent read the trace.

Each cell of Figure~\ref{fig:bycategory} averages 10 to 56 encounters, so part of the spread within a row is sampling noise. To separate it from agent-specific difficulty, we write each score as the encounter's consensus (its mean over agents) plus the agent's deviation from it, and shuffle the deviations across encounters; this keeps each category's difficulty and each agent's overall level and removes any agent-specific category effect. Against 2{,}000 such permutations, the agent-by-category interaction is significant ($p < 0.001$), but it is modest. The agents' category rankings agree with a mean pairwise Spearman $\rho$ of 0.35, just below the 0.37--0.52 that noise alone produces, so weak agreement in rank is mostly a consequence of small cells. Only two categories vary across agents by more than the 95th percentile of the permuted spread: NSTEMI, from 2.28 to 3.78 (1.50 against 1.16), and pericardial disease, from 1.92 to 3.46 (1.54 against 1.25).

\begin{figure*}[tp]
\centering
\includegraphics[width=\textwidth]{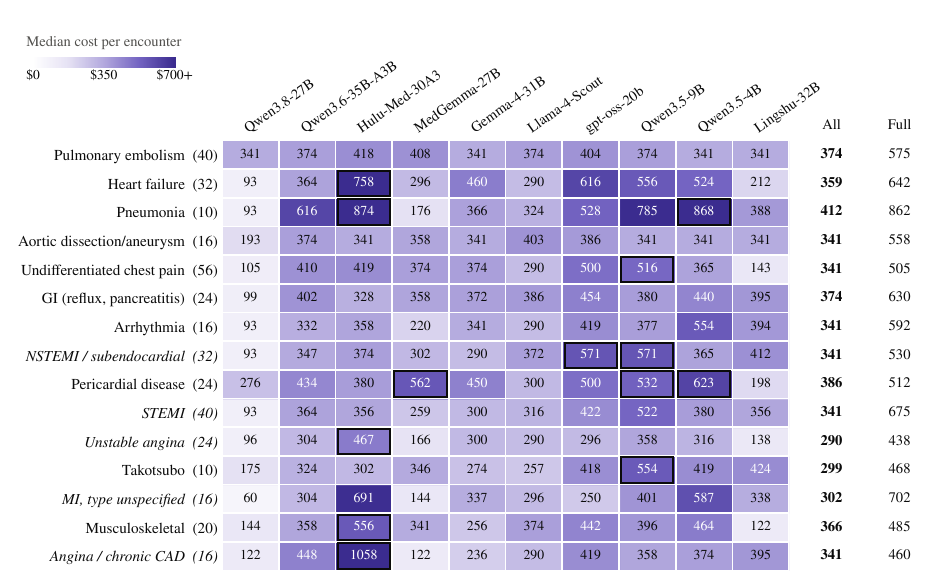}
\caption{Median workup cost per encounter (\$) of each agent by category, gated setting with text-only input; categories, their order and the agents as in Figure~\ref{fig:bycategory}. Outlined cells exceed the category's median full-workup cost, the cost of ordering every study in the record once. \emph{All} is the median over the ten agents' episodes and \emph{Full} the median full-workup cost.}
\label{fig:bycategory-cost}
\end{figure*}

Figure~\ref{fig:bycategory-cost} gives the median cost for the same breakdown. Spending follows the agent more than the case: the pooled median stays between \$290 and \$412 in every category, while the full workup ranges from \$438 to \$862, and most agents' medians sit near \$341, the price of an ECG, one blood draw and a CT angiogram. Qwen3.8-27B is the exception. Its median is \$341 for pulmonary embolism, where it orders the angiogram, and between \$60 and \$276 in every other category, including aortic dissection (\$193), which the same angiogram would settle. Thirteen of the 150 cells exceed the category's full workup: five belong to Hulu-Med-30A3, which repeats orders for results the record does not contain more often than any other agent (1{,}969 such orders over 400 encounters), and four to Qwen3.5-9B.

\paragraph{Counterfactual Analysis} \label{app:counterfactual}

\begin{table}[h]
\centering
\small
\setlength{\tabcolsep}{3pt}
\begin{tabular}{@{}lcccc@{}}
\toprule
 & Unavail. & Util. & Cost & If free \\
\midrule
\multicolumn{5}{@{}l}{\textit{General-purpose}} \\
\quad Qwen3.5-4B & 44.6\% & 47\% & \$389 & \$140 \\
\quad Qwen3.5-9B & 31.3\% & 59\% & \$419 & \$254 \\
\quad Qwen3.8-27B & \textbf{11.4\%} & 52\% & \textbf{\$105} & \textbf{\$93} \\
\quad Qwen3.6-35B-A3B & 20.9\% & 60\% & \$374 & \$257 \\
\quad Gemma-4-31B & 27.3\% & 73\% & \$341 & \$138 \\
\quad Llama-4-Scout & 17.8\% & 62\% & \$328 & \$257 \\
\quad gpt-oss-20b & 32.6\% & 53\% & \$425 & \$183 \\
\addlinespace[2pt]
\multicolumn{5}{@{}l}{\textit{Medical}} \\
\quad MedGemma-27B & 39.7\% & 69\% & \$332 & \$138 \\
\quad Hulu-Med-30A3 & 60.2\% & 53\% & \$422 & \$120 \\
\quad Lingshu-32B & 33.2\% & \textbf{75\%} & \$327 & \$105 \\
\bottomrule
\end{tabular}
\caption{Counterfactual requests in the gated text-only setting. Unavail.: share of an agent's evidence requests that return no result, because the encounter does not contain that study. Util.: share of the evidence channels the encounter does contain that the agent pulled at least once before diagnosing. Cost: median workup cost. If free: the same median when requests returning no result are not charged.}
\label{tab:counterfactual}
\end{table}

The environment charges a request that returns no result in full (\S\ref{sec:task}), which matches billing in a real department but means an agent pays for information it never receives. Table~\ref{tab:counterfactual} reports how much of each agent's workup is spent that way. Costs are recomputed from the transcripts with the price list of Appendix~\ref{app:cost}; the reconstruction reproduces the recorded cost exactly on 395 of 400 episodes.

\paragraph{Unavailable-order rate.} The agents differ by a factor of five, from 11.4\% of requests for Qwen3.8-27B to 60.2\% for Hulu-Med-30A3, and the ordering broadly tracks the gated score: the agents that probe most blindly, Hulu-Med-30A3, Qwen3.5-4B and MedGemma-27B, sit in the lower half of Table~\ref{tab:main}. Anticipating which studies an encounter will contain is part of the task, since a test the treating team never ran returns nothing yet still consumes budget and simulated time.

\paragraph{Available-evidence utilisation.} No agent retrieves most of what is there: utilisation runs from 47\% to 75\%, and the two extremes are instructive. Lingshu-32B pulls the most, 75\%, and scores lowest of the ten; Qwen3.8-27B pulls 52\% and scores highest. Coverage of the record is therefore not what separates the agents, which is consistent with the anchors of dicussion: under the oracle setting the whole record is in context for every agent, and the spread widens rather than closes.

\paragraph{Counterfactual cost sensitivity.} Between 32\% and 84\% of what the agents spend buys nothing. Waiving those charges compresses the costs, from a 4.0$\times$ spread between the cheapest and the dearest agent to 2.8$\times$, and moves seven of the ten positions in the cost ranking. Two moves are large: Hulu-Med-30A3 rises from ninth-cheapest to third, because 83.7\% of its spending goes on studies that do not exist, while Llama-4-Scout falls from third to last, because it probes conservatively and almost everything it orders returns a result. Qwen3.8-27B is cheapest under either rule. The cost column of Table~\ref{tab:main} therefore measures two things at once, how much evidence an agent buys and how well it predicts what is there to buy, and only the first survives the counterfactual.

\FloatBarrier

\begin{figure*}[tp]
\centering
\begin{minipage}[t]{0.495\textwidth}\centering
\includegraphics[width=\linewidth]{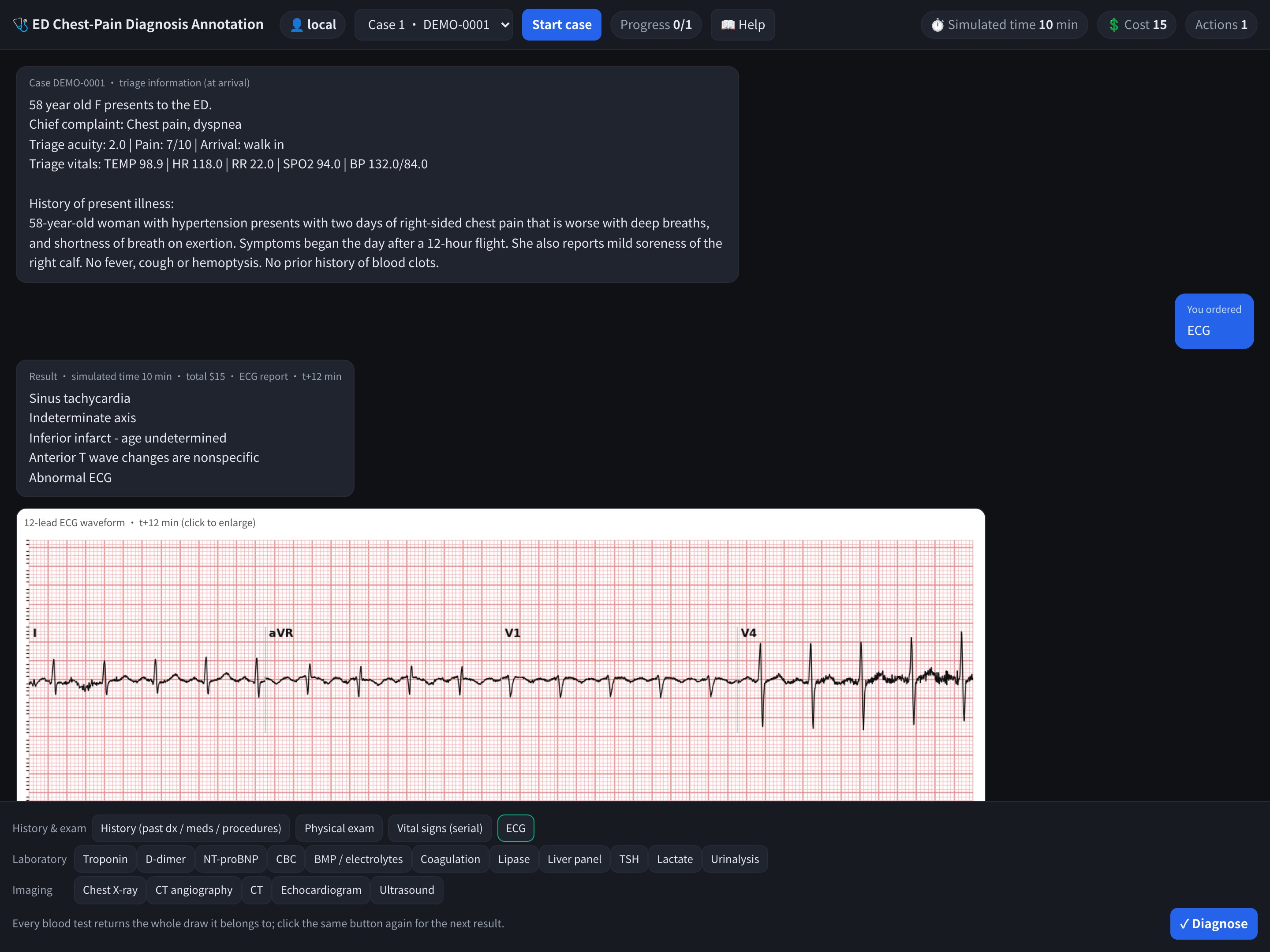}\\[2pt]
{\small (a)}
\end{minipage}\hfill
\begin{minipage}[t]{0.495\textwidth}\centering
\includegraphics[width=\linewidth]{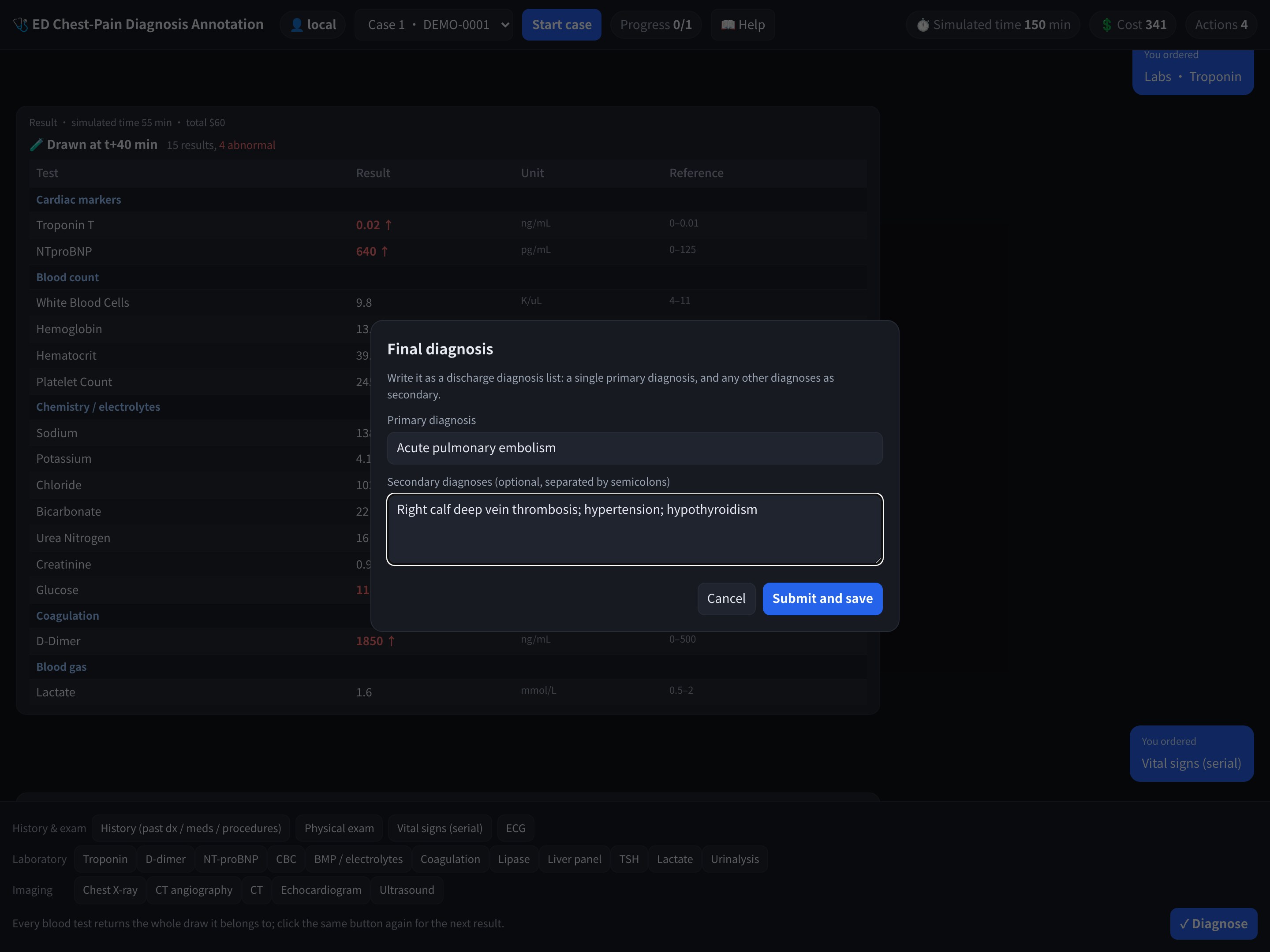}\\[2pt]
{\small (b)}
\end{minipage}
\caption{(a) Triage information and the first action: the ECG returns the machine report and the 12-lead waveform. (b) After a blood draw, serial vital signs and a CT angiogram, the physician enters the final diagnosis. The buttons at the bottom correspond one-to-one to the agent's actions; the header shows the simulated time, cost and number of actions so far.}
\label{fig:annotation-tool}
\end{figure*}

\section{Annotation Tool} \label{app:annotation}

To compare agents with physicians on the same task, we built a web tool in which a physician works up an encounter in the environment the agents face (Figure~\ref{fig:annotation-tool}). The server wraps the same environment object the agents query, so every button corresponds to an agent action with the same price, turnaround time and release rule. The server never loads the reference diagnosis, so it cannot reach the browser. The code will be available on GitHub soon.

\paragraph{Workflow.} An annotation proceeds in four steps.
\begin{enumerate}[leftmargin=1.5em, itemsep=1pt, topsep=2pt]
\item Each physician opens a personal link, which shows only the encounters assigned to them and keeps their progress across sessions. On first use the page shows the instructions: a button can be pressed again for the next result, a blood test returns its whole draw, cost and simulated time accumulate, the time-critical diagnoses must be excluded, and case content must not be copied or shared.
\item An encounter starts from the triage information and the at-arrival history, the same initial observation the agent receives.
\item The physician orders the history, examination, serial vital signs, ECG, one of eleven laboratory panels or one of five imaging studies. Each result appears as a message: laboratory results as a table grouped by category with abnormal values flagged, vital signs as a table with the newest set highlighted, and each ECG as its machine report followed by the 12-lead waveform. The header shows the running cost, simulated time and number of actions.
\item The physician submits a primary diagnosis and optional secondary diagnoses, in the discharge-list format the agents use. Until the next encounter is opened, the diagnosis can be retracted and the workup continued.
\end{enumerate}
Each submission is stored with the diagnosis, the full action log, the cost, the simulated time and any retracted diagnoses, so a physician's episode can be graded by the same judge and priced by the same cost model as an agent's. Encounters flagged in pre-review as lacking the evidence their discharge diagnosis requires carry a folded note; since opening it reveals the diagnosis, whether it was opened is recorded with the submission.

\paragraph{Language.} The interface has a Chinese and an English version. The Chinese version, used by our physicians, shows the case text translated by a locally served model, keeps the original US units and adds conversions to SI units. The English version shows the case exactly as the agents receive it. The environment, prices and clock are the same in both.

\section{Physician Evaluation} \label{app:physicians}

To place the agents against clinical practice, four attending physicians from the emergency department of the same hospital completed the task through the annotation tool (Appendix~\ref{app:annotation}), using its Chinese interface. The 400 encounters were randomly divided into four disjoint sets of 100, one per physician, so every encounter was worked up by exactly one physician. Each physician saw only their own set and received the same initial presentation, actions, prices and step-by-step release of evidence as the agents.

Three physicians completed all 100 of their encounters and one completed 77, for 377 in total. We then apply a quality filter to the annotation process, not to its outcome: an episode is excluded if it contains fewer than two actions or if less than 30 seconds elapsed between opening the encounter and submitting the diagnosis. Thirty seconds is below the time needed to read a returned laboratory panel or radiology report, so such an episode records a submission rather than a workup. The filter removes 70 of the 377 episodes: 19 with no action at all, one with a single action, and 50 completed in under half a minute. They are not spread evenly, falling almost entirely on two annotators, 55 of Physician C's 100 episodes and 14 of Physician A's 77, against one for Physician D and none for Physician B. That the excluded episodes are genuine lapses rather than easy encounters answered at a glance is visible in the agents: on these 70 encounters the ten agents lose 0.43 points relative to their own average, whereas the physicians lose 0.87, and the excluded submissions score 1.80, close to what the agents reach with no evidence at all in the blind anchor. The remaining 307 encounters are the usable set.

The physicians wrote their diagnoses in Chinese and in the ED style of their hospital, which is terser than the discharge-list format the agents produce. Grading the full list would therefore measure documentation style as much as diagnosis, so each physician diagnosis list is translated into English and graded alone. We found that directly comparing Chinese diagnoses with the groundtruth may reduce the model's ability to follow instructions. On a 25-episode check, translating a Chinese diagnosis before grading raises its score by 0.24 on average, which is why the translation is applied to all of them. Table~\ref{tab:physicians} also reports Qwen3.8-27B on the same 307 encounters, in the multimodal setting, since the annotation tool shows the physicians the same rendered 12-lead ECG.

\begin{table*}[t]
\centering
\small
\begin{tabular*}{\textwidth}{@{\extracolsep{\fill}}lccccccc@{}}
\toprule
 & Encounters & Score $\uparrow$ & Acc. $\uparrow$ & DMR $\downarrow$ & Cost, median (IQR) $\downarrow$ & Imaging $\downarrow$ & Turns \\
\midrule
Physician A & 63 & 2.78\,$\pm$\,1.57 & 33.3\% & \textbf{26.1\%} (23) & \$538 (340--736) & \$447 & 9 \\
\quad\textit{Qwen3.8-27B} & \textit{63} & \textit{2.97\,$\pm$\,1.48} & \textit{46.0\%} & \textit{43.5\% (23)} & \textit{\$105 (60--341)} & \textit{\$33} & \textit{4} \\
\addlinespace[3pt]
Physician B & 100 & 2.82\,$\pm$\,1.43 & 41.0\% & 25.6\% (39) & \$941 (918--946) & \$881 & 10 \\
\quad\textit{Qwen3.8-27B} & \textit{100} & \textit{3.68\,$\pm$\,1.38} & \textit{64.0\%} & \textit{20.5\% (39)} & \textit{\$122 (93--341)} & \textit{\$33} & \textit{4} \\
\addlinespace[3pt]
Physician C & 45 & 2.62\,$\pm$\,1.71 & 42.2\% & 50.0\% (24) & \$803 (446--953) & \$728 & 9 \\
\quad\textit{Qwen3.8-27B} & \textit{45} & \textit{3.51\,$\pm$\,1.46} & \textit{60.0\%} & \textit{20.8\% (24)} & \textit{\$93 (60--329)} & \textit{\$33} & \textit{4} \\
\addlinespace[3pt]
Physician D & 99 & 2.58\,$\pm$\,1.56 & 32.3\% & 35.9\% (39) & \$552 (350--585) & \$478 & 7 \\
\quad\textit{Qwen3.8-27B} & \textit{99} & \textit{3.38\,$\pm$\,1.39} & \textit{55.6\%} & \textit{20.5\% (39)} & \textit{\$105 (60--296)} & \textit{\$33} & \textit{4} \\
\midrule
All physicians & 307 & 2.70\,$\pm$\,1.54 & 36.8\% & 33.6\% (125) & \$613 (380--941) & \$511 & 8 \\
\quad\textit{Qwen3.8-27B} & \textit{307} & \textit{3.41\,$\pm$\,1.43} & \textit{57.0\%} & \textit{24.8\% (125)} & \textit{\$105 (60--338)} & \textit{\$33} & \textit{4} \\
\bottomrule
\end{tabular*}
\caption{Physician results on their usable encounters.}
\label{tab:physicians}
\end{table*}

On the same encounters, the physicians are correct less often than Qwen3.8-27B (36.8\% against 57.0\%) and spend far more to get there, a median of \$613 and eight turns against \$105 and four, but they miss time-critical diagnoses at a comparable rate, 33.6\% against 24.8\%. The four differ from one another more than the agents do: Physician B orders the most, at a median of \$941, and has the highest score and the lowest dangerous-miss rate, while Physician C misses half of the time-critical encounters that survive the filter, a spread of 25.6\% to 50.0\% against 22.4\% to 49.6\% across the nine multimodal agents. These numbers therefore describe four clinicians working through machine-translated records in an unfamiliar interface, and should be read as a reference point rather than as the level a deployed system must reach. Part of the cost gap is a habit of practice rather than a difference in judgement. The physicians order 3.07 imaging studies per encounter, 36\% of all their actions, and image at least once in 93\% of encounters, against 0.98 and 66\% for Qwen3.8-27B. In the Chinese emergency departments they work in, radiography, CT and bedside ultrasound are inexpensive relative to the cost of an admission and are usually available within the same shift, so imaging early is a reasonable default there; priced at the 2025/26 US Medicare rates this benchmark uses, that same habit reads as the most expensive workup in the comparison. Whether an agent outperforms a physician also depends on which physician: Physicians A and B hold their dangerous-miss rate at 26.1\% and 25.6\%, close to the 22.4\% of the best agent and ahead of five of the nine, and on A's own encounters the agent misses 43.5\% against A's 26.1\%, while Physician C misses half. The agents therefore clear some of these clinicians on safety and not others. We did not record years of experience, so the spread cannot be attributed to seniority from these data alone.

\section{EDBench-CP Prompts} \label{app:prompts}

This appendix lists the prompts of the evaluation pipeline; the prompts for the at-arrival history rewrite and for the judge are given in Figures~\ref{fig:hpi-prompt} and~\ref{fig:judge-prompt}. Figure~\ref{fig:agent-prompt} gives the agent's system prompt in the gated setting, Figure~\ref{fig:episode-msgs} the fixed messages the environment sends during an episode, and Figure~\ref{fig:anchor-prompt} the prompt of the blind and oracle anchors. All prompts are reproduced with their sections labelled.

\begin{figure*}[tp]
\begin{promptbox}{Agent System Prompt (Gated Setting)}
\pl{Role.} You are an emergency physician working up a patient in the ED. You begin with only the triage presentation and reveal information ONE action at a time. Each action costs money and takes time, so reach the correct principal diagnosis as efficiently and safely as you can.\\[2pt]
\pl{Good practice.}
\begin{pitems}
\item Some diagnoses are low-probability but time-critical and must be actively excluded before discharge: STEMI/ACS, pulmonary embolism, aortic dissection, pneumothorax.
\item Don't order everything. Prefer the cheapest test that best separates your leading differentials.
\item Beware confounders (e.g.\ baseline troponin elevation in renal failure); don't over-call an MI on a mildly abnormal value.
\item ``Undifferentiated / non-cardiac chest pain'' is a legitimate final answer once the dangerous causes are reasonably excluded.
\end{pitems}
\pl{How to reply.} Write a short reasoning in plain prose, then end with EITHER one action line, OR your final diagnosis block.\\[2pt]
\pl{Actions.}
\begin{pitems}
\item \texttt{ACTION: history} (past dx, meds, prior procedures)
\item \texttt{ACTION: exam} (physical exam)
\item \texttt{ACTION: vitals} (serial vital signs so far)
\item \texttt{ACTION: ecg} (12-lead ECG report)
\item \texttt{ACTION: labs <panel>}, panel: troponin | d\_dimer | bnp | cbc | bmp | coags | lipase | lft | tsh | lactate | urine. Ordering ANY panel costs the same and returns the WHOLE blood draw: every analyte resulted from that same sample, not just the named test. Panel choice only affects WHICH draw/timepoint you target and its turnaround time, not price. `urine' is the EXCEPTION: urinalysis is a separate specimen, cheaper than a blood draw, and returns only the urine dipstick/microscopy result.
\item \texttt{ACTION: imaging <modality>}, modality: cxr | cta | ct | echo | us
\end{pitems}
\pl{ECG with images} (multimodal setting only; replaces the \texttt{ecg} line above): 12-lead ECG; you receive the ACTUAL TRACE as an image, at 25\,mm/s and 10\,mm/mV on standard grid, alongside the machine-generated report. Read the waveform yourself: rate, rhythm, axis, intervals, ST segments and T waves lead by lead. Where the trace and the machine report disagree, trust the trace and say so.\\[2pt]
\pl{Final diagnosis.} When you are ready to conclude, give your FINAL DIAGNOSIS as a discharge-style list of ALL the diagnoses you believe this encounter has (a single Primary diagnosis plus any Secondary diagnoses), starting with a line `DIAGNOSE:', for example:\par\smallskip
{\ttfamily DIAGNOSE:\\
Primary: acute pulmonary embolism\\
Secondary: paroxysmal atrial fibrillation; hypertension; type 2 diabetes}\par\smallskip
\pl{Rules.} An action reply MUST be a single line starting `ACTION:'. A diagnosis reply MUST start with a line `DIAGNOSE:' and then list the diagnoses. Do not do both in one reply.
\end{promptbox}
\caption{System prompt of the agent in the gated setting. In the multimodal setting the description of the \texttt{ecg} action is replaced as shown; everything else is identical.}
\label{fig:agent-prompt}
\end{figure*}

\begin{figure*}[tp]
\begin{promptbox}{Episode Messages}
\raggedright
\pl{First user turn.} {\ttfamily INITIAL PRESENTATION (t=0):} followed by the presentation and {\ttfamily Take your first action.} The presentation reads:\par\smallskip
{\ttfamily \{age\} year old \{sex\} presents to the ED.\\
Chief complaint: \{chief complaint\}\\
Triage acuity: \{acuity\} | Pain: \{pain\}/10 | Arrival: \{arrival\}\\
Triage vitals: TEMP \{t\} | HR \{hr\} | RR \{rr\} | SPO2 \{spo2\} | BP \{sbp\}/\{dbp\}\\
History of present illness: \{at-arrival history\}}\par\smallskip
\pl{After each valid action.} {\ttfamily (t=\{clock\}min, cost=\$\{cost\}) Observation: \{observation\} Next action.} In the multimodal setting the ECG image is attached to the observation that releases it.\\[2pt]
\pl{Environment replies.}
\begin{pitems}
\item No further result of the requested kind: {\ttfamily Ordered --- no further this test result available (t=\{clock\}min).} (``this study'' for imaging)
\item Unrecognised action: {\ttfamily Unknown action `\{action\}'. Valid: history, exam, vitals, ecg, labs(panel), imaging(modality), diagnose(text).}
\item Reply without an action or a diagnosis: {\ttfamily Your reply must end with a line starting `ACTION:' or `DIAGNOSE:'. Try again.}
\item Last step (step limit reached): {\ttfamily STOP: budget/step limit reached. Give your FINAL DIAGNOSIS now, starting with a `DIAGNOSE:' line, listing Primary and any Secondary diagnoses.}
\end{pitems}
\end{promptbox}
\caption{Fixed messages the environment sends during an episode. Values in braces are filled in per encounter and step.}
\label{fig:episode-msgs}
\end{figure*}

\begin{figure*}[tp]
\begin{promptbox}{Anchor Prompt (Blind and Oracle)}
\pl{Role.} You are an emergency physician. You are given a chest-pain encounter and must give your FINAL DIAGNOSIS as a discharge-style list: a single Primary diagnosis plus any Secondary diagnoses.\\[2pt]
\pl{Instruction.} Reply with AT MOST TWO SENTENCES of reasoning, then a line `DIAGNOSE:' followed by the list, e.g.\par\smallskip
{\ttfamily DIAGNOSE:\\
Primary: acute pulmonary embolism\\
Secondary: paroxysmal atrial fibrillation; hypertension}\par\smallskip
\pl{Rules.}
\begin{pitems}
\item ``Undifferentiated / non-cardiac chest pain'' is a legitimate answer when no acute cause is apparent.
\item The `DIAGNOSE:' line is mandatory. Do not think out loud at length: measured on the oracle prompt, one model returned 15,000 characters of deliberation and never emitted the block, which makes the answer ungradable.
\end{pitems}
\pl{Input.} The presentation, in the same format as in the gated setting, followed by
\begin{pitems}
\item \emph{Blind:} This is all that is known at triage. No tests have been done and none are available. Commit to your best diagnosis now.
\item \emph{Oracle:} Everything below was obtained during this encounter. Nothing is withheld: you do not need to order anything. This is followed by every evidence item of the encounter (history, examination, serial vital signs, ECGs, laboratory results and imaging reports), in the format the agent receives them in the gated setting.
\end{pitems}
\end{promptbox}
\caption{Prompt of the single-turn blind and oracle anchors.}
\label{fig:anchor-prompt}
\end{figure*}

\section{Case Studies} \label{app:case}
\subsection{Text-only Case}
Figure~\ref{fig:case} follows one complete episode: Qwen3.8-27B on the pulmonary-embolism encounter of Appendix~\ref{app:example}, with text-only input. The agent's turns are reproduced verbatim, including its reasoning before each action; the environment's replies are summarised to the findings they contain. The agent starts from the most dangerous cardiac explanation, uses each result to update its differential, and moves to pulmonary embolism as soon as the first blood draw makes it likely. It reaches the correct diagnosis in four steps for \$341.

\begin{figure*}[tp]
\begin{casebox}{Case Study: A Complete Episode (Qwen3.8-27B), 1/2}
\begin{presbox}
A 55-year-old man walks into the ED with chest pain, triage acuity 2 and pain 10/10. Heart rate 124, blood pressure 166/98, respiratory rate 20, oxygen saturation 98\%, temperature 98.4\,\textdegree F. For three weeks he has had substernal chest pain that comes on after walking about a block, worsens with breathing and resolves with rest, together with shortness of breath. He has no prior DVT or PE, no recent travel, immobility or trauma, and chronic neck and lower back pain; his primary care physician sent him to the ED.\par\smallskip
{\sffamily\scriptsize\color{caseframe}Presentation at arrival, the only information available at the start of the episode (summarised).}
\end{presbox}

\caseturn{1}
\begin{agentturn}
\caserole{agentframe}{AGENT}
The presentation is concerning for acute coronary syndrome given exertional substernal chest pain with dyspnea, tachycardia, and hypertension; a 12-lead ECG is the fastest, highest-yield initial test to exclude STEMI/ACS.\par\smallskip
\texttt{ACTION: ecg}
\end{agentturn}
\begin{envturn}
\caserole{envframe}{ENVIRONMENT \hfill \$15, running total \$15}
12-lead ECG, machine report: sinus tachycardia, indeterminate axis, inferior infarct of undetermined age, nonspecific anterior T-wave changes. No ST elevation is reported.\par\smallskip
\centering\includegraphics[width=0.8\linewidth]{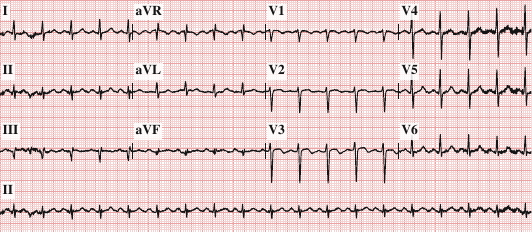}\par\smallskip
{\sffamily\scriptsize\color{envframe!70!black}The trace behind the report (25\,mm/s, 10\,mm/mV). With text-only input the agent receives the report; with image input it also receives the trace.}
\end{envturn}
\casenotes{The agent opens with the cheapest test that can confirm the most time-critical diagnosis. Without ST elevation, STEMI becomes unlikely, but NSTEMI and pulmonary embolism remain open.}

\caseturn{2}
\begin{agentturn}
\caserole{agentframe}{AGENT}
The ECG shows nonspecific changes and an old inferior infarct, but no acute ST elevation. Given the classic exertional substernal chest pain with dyspnea, tachycardia, and hypertension, ACS (NSTEMI/unstable angina) is the leading concern. A troponin is the most critical next step to evaluate for myocardial injury.\par\smallskip
\texttt{ACTION: labs troponin}
\end{agentturn}
\begin{envturn}
\caserole{envframe}{ENVIRONMENT \hfill \$45, running total \$60}
First blood draw, 36 results. Troponin T below the detection limit ($<$0.01\,ng/mL). D-dimer 2{,}062\,ng/mL (reference 0--500), flagged abnormal. Mildly abnormal red-cell indices (MCV 101\,fL) and neutrophils 74\%. Electrolytes, renal and liver function, lipase, glucose and blood counts otherwise within reference ranges.
\end{envturn}
\casenotes{The agent asked for troponin to test ACS, but the request returns the whole blood draw, so the same result also contains the D-dimer. A negative troponin argues against infarction, while the raised D-dimer points elsewhere.}
\end{casebox}
\caption{One complete episode of Qwen3.8-27B, turns 1--2 (continued in Figure~\ref{fig:case2}). Agent turns are verbatim; environment replies are summarised.}
\label{fig:case}
\end{figure*}

\begin{figure*}[tp]
\begin{casebox}{Case Study: A Complete Episode (Qwen3.8-27B), 2/2}
\caseturn{3}
\begin{agentturn}
\caserole{agentframe}{AGENT}
The troponin is negative, which is reassuring against acute MI. However, the D-dimer is significantly elevated at 2062 ng/mL (ref $<$500). Combined with pleuritic chest pain, tachycardia (HR 124), and dyspnea, pulmonary embolism is now the leading concern and must be actively excluded. The ECG's sinus tachycardia is also consistent with PE. CTA of the chest is the definitive next step.\par\smallskip
\texttt{ACTION: imaging cta}
\end{agentturn}
\begin{envturn}
\caserole{envframe}{ENVIRONMENT \hfill \$281, running total \$341}
CT angiography of the chest: filling defects in the pulmonary arteries of all lobes, the largest in the distal right main pulmonary artery. The main pulmonary artery is mildly enlarged (37\,mm), with septal flattening and right ventricular prominence suggesting right heart strain. The aorta is normal without acute pathology; no pericardial or pleural effusion; a small hiatal hernia; several pulmonary nodules under 4\,mm.
\end{envturn}
\casenotes{After one negative troponin the agent leaves the coronary hypothesis instead of repeating the troponin, the opposite of the anchoring on acute coronary syndrome described in \S\ref{sec:discussion}. The CT angiogram confirms pulmonary embolism and, in the same study, excludes aortic dissection.}

\caseturn{4}
\begin{agentturn}
\caserole{agentframe}{AGENT}
\texttt{DIAGNOSE:}\\
\texttt{Primary: acute pulmonary embolism with right heart strain}\\
\texttt{Secondary: hypertension; incidental pulmonary nodules; hiatal hernia; chronic neck and back pain}
\end{agentturn}
\casenotes{The agent commits once the imaging confirms its leading hypothesis, without ordering the echocardiogram or a repeat troponin. Its reasoning for the diagnosis is the one given in turn~3; the final reply contains only the list. The secondary diagnoses collect the incidental findings of the CT angiogram and the history.}

\begin{presbox}
\textbf{Reference} (seen only by the judge): pulmonary embolism, unprovoked.\\
\textbf{Judge} (score 5): ``The physician correctly named acute pulmonary embolism (with right heart strain) as the primary diagnosis, matching the ground truth, and the secondary items are plausible and harmless, though they add little relevant coverage.''\\
\textbf{Episode}: 4 steps, \$341; correct diagnosis.
\end{presbox}
\end{casebox}
\caption{Continuation of Figure~\ref{fig:case}: turns 3--4 and the outcome of the episode.}
\label{fig:case2}
\end{figure*}

\subsection{Image \& Text Case}
Figures~\ref{fig:case3}--\ref{fig:case7} follow a longer episode with image input: Qwen3.6-35B-A3B on an encounter whose discharge diagnosis is a descending thoracic aortic dissection, one of the time-critical diagnoses. The ECG's machine report suggests a possible old anteroseptal infarct. Reading the trace, the agent sets that report aside, works up pulmonary embolism, and finds the dissection on the CT angiogram it orders for embolism. It is graded correct after seven steps and \$655. With text-only input, the same model on the same encounter follows the infarct report, orders troponin twice, never images the chest beyond a radiograph, and diagnoses musculoskeletal pain, a dangerous miss.

\begin{figure*}[tp]
\begin{casebox}{Case Study: A Longer Episode with Image Input (Qwen3.6-35B-A3B), 1/5}
\begin{presbox}
A 62-year-old man walks into the ED with right-sided chest pain, triage acuity 2 and pain 10/10. Heart rate 79, blood pressure 126/81, respiratory rate 18, oxygen saturation 100\%, temperature 98.4\,\textdegree F. He has had constant, sharp right-sided chest pain since the previous evening, without shortness of breath. He has no past medical history; a similar episode some years earlier resolved after ED observation and a negative stress test.\par\smallskip
{\sffamily\scriptsize\color{caseframe}Presentation at arrival (summarised).}
\end{presbox}

\caseturn{1}
\begin{agentturn}
\caserole{agentframe}{AGENT}
The patient presents with acute, sharp, right-sided chest pain. While the description (sharp, pleuritic-sounding, right-sided) and history (prior negative workup) might suggest a benign etiology like musculoskeletal pain or costochondritis, the acuity (Pain 10/10) and age (62) mandate the exclusion of life-threatening causes such as Acute Coronary Syndrome (ACS), Pulmonary Embolism (PE), and Aortic Dissection.\par\smallskip
The most efficient first step is to obtain an ECG. It is low-cost, rapid, and critical for ruling out STEMI or other ischemic changes. Even with right-sided pain, atypical presentations of ACS occur, and baseline data is needed before ordering more expensive tests like CTA or labs.\par\smallskip
\texttt{ACTION: ecg}
\end{agentturn}
\begin{envturn}
\caserole{envframe}{ENVIRONMENT \hfill \$15, running total \$15}
12-lead ECG, machine report: \texttt{Sinus bradycardia with sinus arrhythmia. Possible anteroseptal infarct - age undetermined. Abnormal ECG.}\par\smallskip
\centering\includegraphics[width=0.8\linewidth]{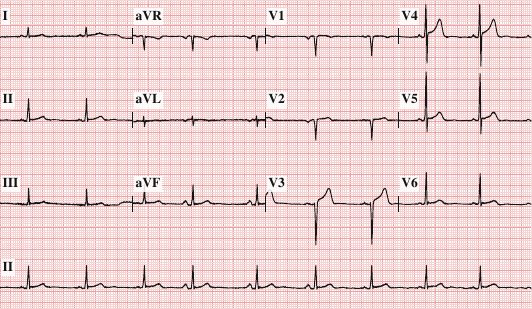}\par\smallskip
{\sffamily\scriptsize\color{envframe!70!black}The trace sent to the agent with the report (25\,mm/s, 10\,mm/mV).}
\end{envturn}
\casenotes{The agent lists aortic dissection among the causes to exclude, but none of its later tests is chosen for it. In the text-only run, the report's ``possible anteroseptal infarct'' is all the agent learns from the ECG, and it spends the rest of that episode excluding an infarct.}
\end{casebox}
\caption{A longer episode: Qwen3.6-35B-A3B with image input, turn 1 (continued in Figures~\ref{fig:case4}--\ref{fig:case7}). Agent turns are verbatim; environment replies other than the ECG report are summarised.}
\label{fig:case3}
\end{figure*}

\begin{figure*}[tp]
\begin{casebox}{Case Study: A Longer Episode with Image Input (Qwen3.6-35B-A3B), 2/5}
\caseturn{2}
\begin{agentturn}
\caserole{agentframe}{AGENT}
The ECG shows sinus bradycardia with sinus arrhythmia. The machine report suggests ``possible anteroseptal infarct,'' but looking at the trace:
\begin{pitems}
\item There are no significant ST-segment elevations in V1-V4.
\item V1-V3 show deep S waves, which can be normal or suggest RVH, but there is no acute ST elevation.
\item V4-V6 show upright T waves and no significant ST depression.
\item The ``sinus arrhythmia'' is a benign finding, often respiratory in origin.
\item The ``possible infarct'' comment is likely a false positive due to the deep S waves in V1-V3 mimicking Q waves or just algorithmic overcall.
\end{pitems}
Given the patient's pain is sharp, right-sided, and pleuritic-sounding (implied by ``sharp''), and the ECG is essentially non-ischemic (no STEMI), ACS is less likely but not fully excluded without troponin. However, the next most dangerous differential to consider given the ``sharp'' pain and potential pleuritic nature is Pulmonary Embolism (PE) or Pneumothorax. A CXR is a good next step to rule out pneumothorax, pneumonia, or other lung pathology, and it is cheaper than a CTA. It also provides a baseline for heart size and lung fields.\par\smallskip
Alternatively, I could order labs (Troponin) to rule out NSTEMI. However, with a non-diagnostic ECG and a pain description that is not classic for cardiac ischemia (sharp, right-sided), the yield of immediate troponin is lower than getting a CXR to look for other causes. Also, if the CXR is normal, I can then proceed to PE workup (D-dimer or CTA) or cardiac workup (troponin) based on risk.\par\smallskip
Let's get a CXR to rule out pneumothorax and other lung issues.\par\smallskip
\texttt{ACTION: imaging cxr}
\end{agentturn}
\begin{envturn}
\caserole{envframe}{ENVIRONMENT \hfill \$33, running total \$48}
Chest radiograph, PA and lateral: clear, well-expanded lungs without consolidation, effusion or pneumothorax; normal heart and mediastinal contours; intact bones. No acute findings.
\end{envturn}
\casenotes{The agent reads the trace lead by lead and treats the infarct call as an artefact of the small anteroseptal complexes. The encounter's second ECG, about eight and a half hours later, repeats the anteroseptal-infarct read, so this is the agent's own judgement of the trace. Its effect is to move the workup away from the coronary hypothesis.}
\end{casebox}
\caption{Continuation of Figure~\ref{fig:case3}: turn 2.}
\label{fig:case4}
\end{figure*}

\begin{figure*}[tp]
\begin{casebox}{Case Study: A Longer Episode with Image Input (Qwen3.6-35B-A3B), 3/5}
\caseturn{3}
\begin{agentturn}
\caserole{agentframe}{AGENT}
The CXR is normal, ruling out pneumothorax, pneumonia, and significant mediastinal widening (making aortic dissection less likely, though not impossible). The ECG was non-ischemic.\par\smallskip
The patient has sharp, right-sided chest pain. The differential now includes:
\begin{enumerate}[leftmargin=1.5em, itemsep=0pt, parsep=0pt, topsep=1pt]
\item \textbf{Musculoskeletal/Costochondritis:} Very common, fits the ``sharp'' description.
\item \textbf{Pulmonary Embolism (PE):} Must be excluded given the age and acute onset, even with normal vitals and O2 sat.
\item \textbf{Acute Coronary Syndrome (ACS):} Less likely given the pain character and ECG, but still possible.
\item \textbf{GERD/Esophageal spasm:} Possible.
\end{enumerate}
Given the ``sharp'' nature and normal CXR/ECG, PE is a significant concern that needs to be addressed. However, the patient has no risk factors mentioned (no PMH, no recent surgery, no immobilization mentioned). The Wells Score for PE would be low (0 points for clinical signs of DVT, 0 for PE being \#1 diagnosis, 0 for HR$>$100, 0 for surgery/immobilization, 0 for prior DVT/PE, 0 for hemoptysis, 0 for malignancy). The score is likely 0 or 1 (if we count ``alternative diagnosis less likely than PE'' as 3, but here MSK is very likely).\par\smallskip
If Wells is low, a D-dimer is the appropriate next step. If D-dimer is negative, PE is ruled out. If positive, CTA is needed.\par\smallskip
Alternatively, I could order Troponin to rule out NSTEMI. Given the pain is 10/10, this is also prudent.\par\smallskip
Let's look at the cost/efficiency. D-dimer is cheap and rules out PE if negative. Troponin rules out MI.\par\smallskip
Given the pain is ``sharp'' and ``right-sided,'' PE is a strong contender to exclude. MSK is also strong.\par\smallskip
I will order a D-dimer to evaluate for PE. If negative, I can focus on MSK or other benign causes. If positive, I'll need a CTA.\par\smallskip
\texttt{ACTION: labs d\_dimer}
\end{agentturn}
\begin{envturn}
\caserole{envframe}{ENVIRONMENT \hfill \$45, running total \$93}
Blood draw at 49 minutes, one result: D-dimer 1{,}043\,ng/mL (reference 0--500), flagged abnormal.
\end{envturn}
\casenotes{The agent applies a pre-test probability rule (Wells) before choosing between D-dimer and CT angiography, and keeps troponin as an alternative it has not excluded. The panel name selects the blood draw: the D-dimer was measured on a separate draw from the troponin, so this request returns only the D-dimer. Both troponin requests of the text-only run return the other draws, and that run never sees the D-dimer.}

\caseturn{4}
\begin{agentturn}
\caserole{agentframe}{AGENT}
The D-dimer is elevated (1043 ng/mL). In a 62-year-old male, this raises the suspicion for Pulmonary Embolism (PE), although D-dimer is non-specific and can be elevated in many conditions (infection, inflammation, age-related).\par\smallskip
Given the elevated D-dimer and the patient's presentation of acute chest pain, PE must be actively excluded. The next step is a CT Angiogram (CTA) of the chest to visualize the pulmonary arteries.\par\smallskip
\texttt{ACTION: imaging cta}
\end{agentturn}
\begin{envturn}
\caserole{envframe}{ENVIRONMENT \hfill \$281, running total \$374}
CT angiography of the chest: no filling defect in the pulmonary arteries down to the subsegmental level. A small linear defect in the anterior aspect of the descending aorta, consistent with an intimal flap; the aorta is of normal calibre without intramural hematoma. Bilateral centrilobular emphysema, a 2\,mm right middle lobe nodule, and incompletely characterised liver hypodensities.
\end{envturn}
\casenotes{The CT angiogram is ordered to exclude pulmonary embolism. The same study images the thoracic aorta, so it finds the dissection that no test so far was chosen to detect.}
\end{casebox}
\caption{Continuation of Figure~\ref{fig:case4}: turns 3--4.}
\label{fig:case5}
\end{figure*}

\begin{figure*}[tp]
\begin{casebox}{Case Study: A Longer Episode with Image Input (Qwen3.6-35B-A3B), 4/5}
\caseturn{5}
\begin{agentturn}
\caserole{agentframe}{AGENT}
The CTA rules out Pulmonary Embolism. However, it reveals a ``small linear defect in the anterior aspect of the descending aorta compatible with an intimal flap.'' This is a critical finding suggestive of \textbf{Aortic Dissection} (likely Stanford Type B, given the descending location, though the extent needs clarification).\par\smallskip
Although the patient's BP is currently normal (126/81) and pain is right-sided (atypical for dissection, which is often tearing and central/back), the imaging finding is highly specific. An intimal flap is the hallmark of dissection. This is a life-threatening condition that requires immediate confirmation and management (blood pressure control, surgical consultation).\par\smallskip
The next step is to further characterize the aorta to determine the extent of the dissection (involvement of arch, abdominal aorta, branch vessels) and confirm the diagnosis. A dedicated \textbf{CTA of the Aorta} (or a CT Angiogram of the Chest/Abdomen/Pelvis) is the standard next step. Since the current scan was a ``Non-Coronary CTA Chest,'' it may not have covered the entire aorta optimally for dissection staging. However, often a ``CTA Chest'' with wide coverage can be sufficient if it includes the abdomen. The report says ``CT of the Thorax... visualized solid organs... unremarkable,'' implying limited abdominal view.\par\smallskip
To be safe and efficient, I should order a \textbf{CTA of the Aorta} (or CT Chest/Abdomen/Pelvis with aortic protocol) to define the extent. Alternatively, an \textbf{Echo} (TTE/TEE) can be used to visualize the ascending aorta and pericardium, but CTA is generally superior for defining the full extent of Type B dissections. Given the finding is in the \emph{descending} aorta, TTE might miss it. TEE is invasive. CTA is the best next step.\par\smallskip
Wait, the current CTA already saw the flap. The question is whether I need more imaging or if I can proceed to diagnosis/management. The finding is ``compatible with an intimal flap.'' In the context of acute chest pain, this is a dissection until proven otherwise. I need to know if it extends proximally or distally. The current scan was ``suprasternal notch to upper abdomen.'' This usually covers the thoracic aorta well. The report says ``descending aorta... anterior aspect.''\par\smallskip
If the dissection is limited to the descending aorta (Type B), management is medical (BP control). If it involves the ascending aorta (Type A), it's surgical. The report says ``descending aorta.'' It does not mention the ascending aorta having a flap. The ``CTA Thorax'' section says ``aorta and main thoracic vessels are well opacified... no intramural hematoma.'' This suggests the ascending aorta is clear.\par\smallskip
So, we likely have a \textbf{Type B Aortic Dissection}.\par\smallskip
However, I must be careful. Is this a real dissection or an artifact? The report says ``compatible with.'' In an emergency setting, this is treated as a dissection.\par\smallskip
Do I need another test? The current CTA seems to have covered the necessary area (chest to upper abdomen). It visualized the descending aorta. It did not see the ascending aorta involved.\par\smallskip
I should probably get a \textbf{CTA of the Aorta} (full length) if the current one didn't cover the abdominal aorta fully, or just accept the current findings if they are sufficient. The report says ``CT of the Thorax... visualized solid organs... unremarkable.'' It doesn't explicitly say the abdominal aorta was fully evaluated for dissection.\par\smallskip
Actually, the most critical next step is to \textbf{confirm} this finding and assess for complications. A \textbf{CTA of the Aorta} (Chest/Abdomen/Pelvis) is the gold standard. The current scan was a ``Non-Coronary CTA Chest'' which might have been timed for pulmonary arteries, not necessarily optimal for aortic wall visualization, although it did see the flap.\par\smallskip
Let's order a \textbf{CTA of the Aorta} to fully characterize the dissection extent.\par\smallskip
\texttt{ACTION: imaging cta}
\end{agentturn}
\begin{envturn}
\caserole{envframe}{ENVIRONMENT \hfill \$281, running total \$655}
The record holds no further CT angiogram; the environment replies that no further result of this study is available.
\end{envturn}
\casenotes{The agent recognises the dissection at once, then deliberates over whether the first scan already answers the remaining question. It concludes twice that it does and orders a second CT angiogram anyway. The repeat order returns nothing and adds \$281, 43\% of the episode's cost.}
\end{casebox}
\caption{Continuation of Figure~\ref{fig:case5}: turn 5.}
\label{fig:case6}
\end{figure*}

\begin{figure*}[tp]
\begin{casebox}{Case Study: A Longer Episode with Image Input (Qwen3.6-35B-A3B), 5/5}
\caseturn{6}
\begin{agentturn}
\caserole{agentframe}{AGENT}
The previous CTA Chest identified an intimal flap in the descending aorta, highly suggestive of a Type B Aortic Dissection. The CTA ruled out PE. The ECG was non-ischemic. The CXR was normal.\par\smallskip
The patient has a confirmed (or highly suspected) Aortic Dissection. This is a medical emergency. The immediate management for Type B dissection (descending aorta) is strict blood pressure and heart rate control to reduce shear stress on the aortic wall (anti-impulse therapy).\par\smallskip
I need to check the patient's current vitals to guide therapy. If the BP is elevated, I need to start IV beta-blockers (e.g., labetalol or esmolol).\par\smallskip
\texttt{ACTION: vitals}
\end{agentturn}
\begin{envturn}
\caserole{envframe}{ENVIRONMENT \hfill \$0, running total \$655}
Serial vital signs, one set at 18 minutes: heart rate 62, respiratory rate 16, oxygen saturation 100\%, blood pressure 118/84, pain 8/10.
\end{envturn}

\caseturn{7}
\begin{agentturn}
\caserole{agentframe}{AGENT}
The patient's blood pressure is 118/84 mmHg and heart rate is 62 bpm. These are within the target range for acute aortic dissection management (SBP $<$ 120 mmHg, HR $<$ 60 bpm). The heart rate is slightly above the ideal $<$60 bpm target, but the BP is well-controlled.\par\smallskip
The diagnosis of \textbf{Acute Aortic Dissection (Stanford Type B)} is established by the CTA finding of an intimal flap in the descending aorta. PE has been ruled out. ACS is unlikely given the non-ischemic ECG and atypical pain, though troponin could be checked to be thorough, it is not the primary driver of immediate management.\par\smallskip
The patient requires admission for blood pressure monitoring and medical management. No further diagnostic imaging is immediately necessary as the extent (descending only) is clear from the CTA.\par\smallskip
\texttt{DIAGNOSE:}\\
\texttt{Primary: Acute Type B Aortic Dissection}\\
\texttt{Secondary: Uncomplicated (no malperfusion, rupture, or refractory pain noted)}
\end{agentturn}
\casenotes{In turns 6 and 7 the agent reasons about management, checking the vital signs against blood-pressure and heart-rate targets for dissection. The benchmark grades only the diagnosis. The final reply gives the reasons for the diagnosis: the intimal flap on the CT angiogram, pulmonary embolism excluded by the same study, and a non-ischaemic ECG with atypical pain against ACS, which it leaves without a troponin.}

\begin{presbox}
\textbf{Reference} (seen only by the judge): descending thoracic aorta dissection; chest pain of unknown etiology.\\
\textbf{Judge} (score 4): ``The physician's primary diagnosis, acute Type B aortic dissection, matches the ground-truth descending thoracic aorta dissection (Type B means descending). The secondary entry only qualifies the dissection as uncomplicated and adds nothing else.''\\
\textbf{Episode}: 7 steps, \$655; correct diagnosis.\\
\textbf{Same encounter, text-only input}: 8 steps (ECG, blood draw, examination, chest X-ray, blood draw, history, examination, diagnosis), \$138; primary diagnosis undifferentiated chest pain, likely musculoskeletal; judge score 1, a dangerous miss.
\end{presbox}
\end{casebox}
\caption{Continuation of Figure~\ref{fig:case6}: turns 6--7, the outcome of the episode, and the outcome of the text-only episode on the same encounter.}
\label{fig:case7}
\end{figure*}

Figures~\ref{fig:case8}--\ref{fig:case10} show a failure of Qwen3.8-27B with text-only input that follows the first failure pattern of \S\ref{sec:discussion}. The encounter's discharge diagnosis is a non-ST-elevation myocardial infarction from demand ischaemia with pulmonary oedema, in a patient on peritoneal dialysis who had missed dialysis. The agent receives a troponin T of 0.39\,ng/mL, 39 times the upper reference limit, attributes it to chronic elevation in end-stage renal disease, and notes that a rise would decide the question, but never repeats the measurement. It later cites a normal CK-MB that was never measured and diagnoses non-cardiac chest pain after five steps and \$93. Given the whole record in one turn (the oracle anchor), the same model diagnoses the infarction.

\begin{figure*}[tp]
\begin{casebox}{Case Study: A Failed Episode (Qwen3.8-27B), 1/3}
\begin{presbox}
A 60-year-old woman arrives by ambulance with chest pain, triage acuity 2 and pain 3/10. Heart rate 70, blood pressure 150/45, respiratory rate 16, oxygen saturation 98\%, temperature 98.1\,\textdegree F. For two days she has had constant right-sided chest pain with a cough occasionally productive of white sputum, and nausea with a few episodes of emesis, recently slightly pink; earlier diarrhoea and dark stools have resolved, and she has had no fever. She has hypertension, insulin-dependent diabetes and end-stage renal disease on peritoneal dialysis.\par\smallskip
{\sffamily\scriptsize\color{caseframe}Presentation at arrival (summarised).}
\end{presbox}

\caseturn{1}
\begin{agentturn}
\caserole{agentframe}{AGENT}
I have a 60-year-old woman with constant right-sided chest pain, cough, and several red-flag features: ESRD on dialysis, hypertension, and a very wide pulse pressure (150/45). Even though the pain is mild and right-sided, I need to immediately exclude acute coronary syndrome and look for any ECG clues to ischemia, pericarditis, or other cardiac causes before pursuing more expensive imaging.\par\smallskip
\texttt{ACTION: ecg}
\end{agentturn}
\begin{envturn}
\caserole{envframe}{ENVIRONMENT \hfill \$15, running total \$15}
12-lead ECG, machine report: \texttt{Sinus rhythm. Possible left ventricular hypertrophy. Inferior/lateral ST-T changes may be due to hypertrophy and/or ischemia. Abnormal ECG.}\par\smallskip
\centering\includegraphics[width=0.8\linewidth]{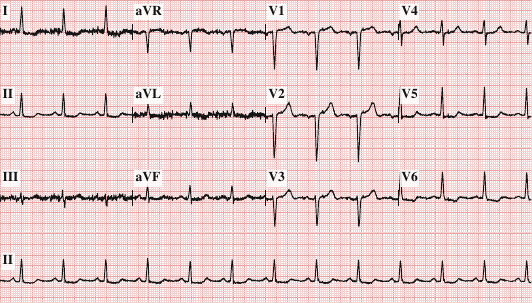}\par\smallskip
{\sffamily\scriptsize\color{envframe!70!black}The trace behind the report (25\,mm/s, 10\,mm/mV). With text-only input the agent receives the report only.}
\end{envturn}
\casenotes{The opening is sound: the agent names the risk factors, notes the wide pulse pressure and starts with the ECG.}

\caseturn{2}
\begin{agentturn}
\caserole{agentframe}{AGENT}
The ECG shows non-specific ST-T changes that could represent LVH or ischemia in a high-risk patient (ESRD, HTN, DM). The wide pulse pressure (150/45) is also concerning. I need to rule out ACS given the ECG findings and risk profile. Troponin is the cheapest and most direct next step to address the possible ischemic changes.\par\smallskip
\texttt{ACTION: labs troponin}
\end{agentturn}
\begin{envturn}
\caserole{envframe}{ENVIRONMENT \hfill \$45, running total \$60}
First blood draw, 33 results. Troponin T 0.39\,ng/mL (reference 0--0.01), flagged abnormal. Urea nitrogen 121\,mg/dL and creatinine 20.5\,mg/dL; anion gap 28\,mEq/L with bicarbonate 21\,mEq/L; haemoglobin 9.5\,g/dL; glucose 175\,mg/dL. Sodium and potassium within reference ranges. The draw contains no CK-MB.
\end{envturn}
\casenotes{The ECG leaves ischaemia open, and the agent orders the test that can settle it.}
\end{casebox}
\caption{A failed episode: Qwen3.8-27B with text-only input, turns 1--2 (continued in Figures~\ref{fig:case9} and~\ref{fig:case10}). Agent turns are verbatim; environment replies other than the ECG report are summarised.}
\label{fig:case8}
\end{figure*}

\begin{figure*}[tp]
\begin{casebox}{Case Study: A Failed Episode (Qwen3.8-27B), 2/3}
\caseturn{3}
\begin{agentturn}
\caserole{agentframe}{AGENT}
We have a troponin T of 0.39 ng/mL, which is markedly elevated (ref 0-0.01). However, the patient has ESRD on peritoneal dialysis. Chronic troponin elevation is common in ESRD, often due to chronic myocardial injury, LVH, and reduced clearance. A single elevated troponin in ESRD is not diagnostic of acute MI without a dynamic rise or clear ischemic symptoms/ECG changes.\par\smallskip
The ECG showed ``Inferior/lateral ST-T changes may be due to hypertrophy and/or ischemia.'' This is non-specific. The pain is right-sided, constant for 2 days, associated with cough and white sputum. This presentation is less typical for ACS (which is usually substernal, pressure-like, radiating). The right-sided chest pain with cough suggests a pulmonary or pleural etiology.\par\smallskip
Other findings:
\begin{pitems}
\item Anion gap 28 (high), Bicarb 21 (low). This is a high anion gap metabolic acidosis. In ESRD, this is expected (uremic acidosis). Urea Nitrogen is 121, Creatinine 20.5.
\item Hemoglobin 9.5, Hct 29.8: Anemia of chronic disease/ESRD.
\item Glucose 175: IDDM.
\item The ``pink'' emesis and ``dark colored stools'' that resolved could suggest a GI bleed, but the Hgb is stable-ish (though we don't have a baseline, 9.5 is typical for ESRD). The dark stools resolved.
\item The wide pulse pressure (150/45) is notable. Could be aortic regurgitation, or just stiff vessels in ESRD/HTN.
\end{pitems}
The main differential now:
\begin{enumerate}[leftmargin=1.5em, itemsep=0pt, parsep=0pt, topsep=1pt]
\item ACS: Troponin is elevated, but likely chronic in ESRD. Need to see if it's rising. But the symptoms (right-sided, 2 days, cough) are atypical.
\item Pulmonary Embolism: Right-sided chest pain, cough. Risk factors? ESRD, maybe immobility? D-dimer would be elevated in ESRD anyway, so not very useful. CTA is the gold standard.
\item Pneumonia/Pleurisy: Cough, white sputum, right-sided pain. CXR would help.
\item Aortic Dissection: Wide pulse pressure, HTN, chest pain. But pain is right-sided and constant for 2 days. Dissection pain is usually tearing and acute.
\item Pericarditis: ESRD patients are prone to uremic pericarditis. Pain is usually pleuritic and positional. ECG might show diffuse ST elevation (not seen here, just ST-T changes).
\end{enumerate}
Given the atypical nature of the pain (right-sided, 2 days, cough) and the confounding troponin elevation from ESRD, I should look for other causes. A chest X-ray is the cheapest and most informative next step to evaluate for pneumonia, pleural effusion, pneumothorax, or aortic widening. It's a standard part of the chest pain workup.\par\smallskip
Let's get a CXR. It's cheap and will help differentiate pulmonary causes (pneumonia, effusion) from cardiac causes (cardiomegaly, aortic widening).\par\smallskip
Wait, the prompt says ``Some diagnoses are low-probability but time-critical and must be actively excluded before discharge: STEMI/ACS, pulmonary embolism, aortic dissection, pneumothorax.''\\
I've done ECG. Troponin is elevated but confounded.\\
CXR is the next logical step to rule out pneumothorax, pneumonia, and look at the aorta.\par\smallskip
Let's do CXR.\par\smallskip
\texttt{ACTION: imaging cxr}
\end{agentturn}
\begin{envturn}
\caserole{envframe}{ENVIRONMENT \hfill \$33, running total \$93}
Chest radiograph, PA and lateral: clear lungs, normal hilar and cardiomediastinal contours, no pneumothorax or effusion, normal pulmonary vascularity. No acute cardiopulmonary process.
\end{envturn}
\casenotes{This is where the episode goes wrong. The agent states the rule that decides the case, that an elevated troponin in renal failure indicates an acute infarction only if it rises, and writes ``Need to see if it's rising,'' but orders a chest radiograph and never repeats the troponin. The agent prompt warns against calling an infarction on a \emph{mildly} abnormal troponin in renal failure; the agent applies that warning to a value 39 times the upper limit. From here the raised troponin is treated as explained, and the pain's atypical features steer the differential toward pulmonary causes.}
\end{casebox}
\caption{Continuation of Figure~\ref{fig:case8}: turn 3.}
\label{fig:case9}
\end{figure*}

\begin{figure*}[tp]
\begin{casebox}{Case Study: A Failed Episode (Qwen3.8-27B), 3/3}
\caseturn{4}
\begin{agentturn}
\caserole{agentframe}{AGENT}
The CXR is clear, ruling out pneumonia, pneumothorax, and gross mediastinal widening. The troponin elevation is likely chronic given ESRD (CK-MB is normal, ECG non-specific, pain atypical for ACS). However, the wide pulse pressure (150/45) is striking and raises concern for aortic regurgitation or dissection. I haven't performed a physical exam yet, which is the cheapest next step and could reveal a murmur, friction rub, or BP differential between arms.\par\smallskip
\texttt{ACTION: exam}
\end{agentturn}
\begin{envturn}
\caserole{envframe}{ENVIRONMENT \hfill \$0, running total \$93}
Admission examination: intubated and sedated; rectal temperature 33.8\,\textdegree C, blood pressure 103/81, pulse 57; pinpoint pupils; right internal jugular central line and urinary catheter in place; clear lungs; regular rhythm without murmurs, rubs or gallops; soft abdomen; cool extremities with oedema, left greater than right.
\end{envturn}
\casenotes{``CK-MB is normal'' has no source: the blood draw contained no CK-MB, and the agent never ordered one. The invented value becomes a second argument for the chronic explanation.}

\caseturn{5}
\begin{agentturn}
\caserole{agentframe}{AGENT}
The exam reveals the patient is now intubated and sedated, which explains the pinpoint pupils and bradycardia. The initial wide pulse pressure (150/45) has resolved to 103/81. No murmurs, rubs, or gallops. Lungs clear. No BP differential mentioned.\par\smallskip
Summary of workup:
\begin{pitems}
\item ECG: Non-specific ST-T changes (LVH vs ischemia) - no STEMI
\item Troponin T: 0.39 ng/mL - elevated but expected in ESRD (chronic myocardial injury, reduced clearance); CK-MB normal; pain atypical for ACS (right-sided, 2 days, with cough)
\item CXR: Clear - no pneumonia, pneumothorax, effusion, or mediastinal widening
\item Exam: No murmur (argues against aortic regurgitation/dissection), no rub (argues against pericarditis), clear lungs
\end{pitems}
The major time-critical diagnoses have been reasonably excluded:
\begin{pitems}
\item ACS/STEMI: No ST elevation, CK-MB normal, atypical pain, troponin likely chronic in ESRD
\item PE: SpO2 98\%, no tachycardia, clear CXR
\item Aortic dissection: Normal mediastinum on CXR, no murmur, no BP differential
\item Pneumothorax: Clear CXR
\end{pitems}
The chest pain is most consistent with a non-cardiac etiology in the setting of ESRD (possibly musculoskeletal, pleuritic, or uremic). The intubation likely occurred for airway protection (given emesis/aspiration risk or altered mental status from uremia).\par\smallskip
\texttt{DIAGNOSE:}\\
\texttt{Primary: Undifferentiated non-cardiac chest pain}\\
\texttt{Secondary: ESRD on peritoneal dialysis; uremic acidosis; hypertension; type 1 diabetes mellitus; anemia of chronic disease; intubation for airway protection}
\end{agentturn}
\casenotes{The environment returns the admission examination, recorded after the patient's ED course: she has been intubated and is hypothermic and bradycardic. The agent explains the intubation as airway protection, does not mention the temperature, and keeps its conclusion. Its final reasoning lists acute coronary syndrome as excluded on four grounds; two of them are the troponin it has declared chronic without a repeat and the CK-MB it never received.}

\begin{presbox}
\textbf{Reference} (seen only by the judge): non-ST-elevation myocardial infarction secondary to demand ischaemia; pulmonary oedema; secondary uraemia from missed dialysis and end-stage renal disease on peritoneal dialysis.\\
\textbf{Judge} (score 1): ``The ground truth chest-pain cause is NSTEMI (demand ischemia) with pulmonary edema, but the physician's primary was undifferentiated non-cardiac chest pain, and NSTEMI, a can't-miss diagnosis, is not named anywhere in the list.''\\
\textbf{Episode}: 5 steps, \$93; wrong diagnosis, a dangerous miss.\\
\textbf{Same encounter, oracle anchor} (whole record in one turn): primary non-ST-elevation myocardial infarction; secondary acute decompensated heart failure, end-stage renal disease, hyperkalaemia and metabolic acidosis; judge score 5.
\end{presbox}
\end{casebox}
\caption{Continuation of Figure~\ref{fig:case9}: turns 4--5, the outcome of the episode, and the oracle anchor on the same encounter.}
\label{fig:case10}
\end{figure*}

\section{The usage of LLMs}
In addition to serving as the research subject, the LLM contributed to this study by polishing the paper (including generating image assets and formatting LaTeX tables and figures) and writing code (including generating annotation tools and auxiliary SQL statements).

\end{document}